\documentclass[final,1p,times]{elsarticle}
\usepackage{colordvi}

\usepackage{epsfig}

\usepackage{amssymb}

\journal{Computer Physics Communications}

\newcommand{\centori}{\texttt{CENTORI}}
\newcommand{\grass}{\texttt{GRASS}}
\newcommand{\vi}{\mathbf{v_i}}
\newcommand{\ve}{\mathbf{v_e}}
\newcommand{\A}{\mathbf{A}}
\newcommand{\B}{\mathbf{B}}

\newcommand{\Aeq}{\mathbf{A}_\mathbf{eq}}
\newcommand{\Beq}{\mathbf{B}_\mathbf{eq}}

\newcommand{\vieq}{\mathbf{v}_{i \; \mathbf{eq}}}
\newcommand{\veeq}{\mathbf{v}_{e \; \mathbf{eq}}}
\newcommand{\E}{\mathbf{E}}
\newcommand{\J}{\mathbf{J}}
\newcommand{\Jeq}{\mathbf{J}_\mathbf{eq}}
\newcommand{\W}{\mathbf{W}}
\newcommand{\q}{\mathbf{q}}
\newcommand{\curl}{\nabla \times}
\newcommand{\X}{\times}

\newcommand{\vistar}{\mathbf{v_i^*}}
\newcommand{\vestar}{\mathbf{v_e^*}}
\newcommand{\Astar}{\mathbf{A^*}}
\newcommand{\Bstar}{\mathbf{B^*}}
\newcommand{\Beqstar}{\mathbf{B}_\mathbf{eq}^*}
\newcommand{\Phistar}{\Phi^*}

\newcommand{\nestar}{n_e^*}
\newcommand{\pistar}{p_i^*}
\newcommand{\pestar}{p_e^*}
\newcommand{\tistar}{T_i^*}
\newcommand{\testar}{T_e^*}
\newcommand{\Wstar}{\mathbf{W^*}}
\newcommand{\Wfstar}{\tilde\Wstar}
\newcommand{\Jstar}{\mathbf{J^*}}
\newcommand{\Jfstar}{\tilde\Jstar}
\newcommand{\Estar}{\mathbf{E^*}}

\newcommand{\va}{v_A}
\newcommand{\nave}{\overline{n_e}}
\newcommand{\Sv}{\mathbf{S_v}}
\newcommand{\Svstar}{\mathbf{S_v^*}}
\newcommand{\Snstar}{S_n^*}

\newcommand{\dd}{\partial}
\newcommand{\dt}{\partial t}
\newcommand{\dpsi}{\partial \psi}
\newcommand{\drho}{\partial \rho}
\newcommand{\dtheta}{\partial \theta}
\newcommand{\dzeta}{\partial \zeta}
\newcommand{\jac}{\mathcal{J}}
\newcommand{\jacstar}{\mathcal{J}^*}

\newcommand{\grad}{\nabla}

\newcommand{\delt}{\Delta t}
\newcommand{\delrho}{\Delta \rho}
\newcommand{\deltheta}{\Delta \theta}

\newcommand{\er}{\mathbf{e_R}}
\newcommand{\ez}{\mathbf{e_Z}}
\newcommand{\ezeta}{\mathbf{e_\zeta}}

\newcommand{\bcovrho}{\mathbf{b^\rho}}
\newcommand{\bcovtheta}{\mathbf{b^\theta}}
\newcommand{\bcovzeta}{\mathbf{b^\zeta}}
\newcommand{\bconrho}{\mathbf{b_\rho}}
\newcommand{\bcontheta}{\mathbf{b_\theta}}
\newcommand{\bconzeta}{\mathbf{b_\zeta}}

\newcommand{\delpsi}{\Delta^* \psi}

\newcommand{\nuparai}{\nu_{i \parallel}}
\newcommand{\nuparae}{\nu_{e \parallel}}

\begin{document}

\begin{frontmatter}



\title{CENTORI: a global toroidal electromagnetic two-fluid plasma turbulence code}


\author[CCFE]{P. J. Knight}  \ead{peter.knight@ccfe.ac.uk}
\author[UnivBristol]{A. Thyagaraja}  \ead{a.thyagaraja@bristol.ac.uk}
\author[HECToR]{T. D. Edwards\fnref{footnote}} \ead{tedwards@cray.com}
\author[UnivEdinburgh,Lund]{J. Hein} \ead{Joachim.Hein@math.lu.se}
\author[CCFE]{M. Romanelli}  \ead{michele.romanelli@ccfe.ac.uk}
\author[CCFE]{K. G. McClements\corref{cor1}} \ead{k.g.mcclements@ccfe.ac.uk}

\cortext[cor1]{Corresponding author.}

\fntext[footnote]{This author's contribution was completed 
while studying as a CASE Ph.D. student at the University of Edinburgh in 
collaboration with CCFE.}

\address[CCFE]{EURATOM/CCFE Fusion Association, Culham Science Centre,
  Abingdon, OX14 3DB, UK}
\address[UnivBristol]{University of Bristol, H. H. Wills Physics Laboratory,
  Bristol BS8 1TL, UK}
\address[HECToR]{Cray Centre of Excellence for HECToR, 2261 JCMB, 
  University of Edinburgh, Edinburgh EH9 3JZ, UK} 
\address[UnivEdinburgh]{EPCC, University of Edinburgh, Mayfield Road, Edinburgh 
  EH9 3JZ, UK}
\address[Lund] {Lunds Universitet, Box 118, 221 00 Lund, Sweden}

\begin{abstract}
A new global two-fluid electromagnetic turbulence code, \centori, has been 
developed for the purpose of studying magnetically-confined fusion plasmas
on energy confinement timescales. This code is used to evolve the combined 
system of electron and ion fluid equations and Maxwell equations in toroidal 
configurations with axisymmetric equilibria. Uniquely, the equilibrium is 
co-evolved with the turbulence, and is thus modified by it.
\centori\/ is applicable to tokamaks of arbitrary aspect ratio 
and high plasma beta. A predictor-corrector, semi-implicit finite difference scheme is
used to compute the time evolution of fluid quantities and fields. Vector 
operations and the evaluation of flux surface averages
are speeded up by choosing the Jacobian of the transformation from
laboratory to plasma coordinates to be a function of the equilibrium 
poloidal magnetic flux. A subroutine, \grass, is used to co-evolve the plasma 
equilibrium by computing the steady-state solutions of a diffusion equation with a
pseudo-time derivative. The code is written in Fortran 95 and is efficiently
parallelized using Message Passing Interface (MPI). 
Illustrative examples of output from simulations of a tearing mode in a large 
aspect ratio tokamak plasma and of turbulence in an elongated conventional 
aspect ratio tokamak plasma are provided.
\end{abstract}


\end{frontmatter}




\section{Introduction}
\label{sec:intro}

Plasma confinement in tokamak experiments is determined partly by binary Coulomb 
collisions between charged particles, but mainly by turbulence and instabilities, which 
occur on scales ranging from particle Larmor radii to the system size. Understanding the
nature of this turbulence is a key goal of thermonuclear fusion research, since the 
confinement time is one of the parameters that must be optimised in order to create burning 
plasma conditions. In order to simulate turbulence in tokamak plasmas it is necessary to 
either average the Vlasov equations of the particle species over gyro-angle (the 
gyrokinetic approach) or take full 
velocity-space moments of these equations (the fluid approach). The lower dimensionality
of fluid models makes it possible to simulate larger systems over longer timescales, and 
for this reason fluid codes continue to play an important role in tokamak plasma 
modelling. Some of these codes are based on electrostatic models \cite{Ottaviani,Garbet} or 
employ flux tube geometry \cite{Scott}, while others are designed specifically for the 
purpose of simulating edge plasma phenomena, such as edge localised modes (ELMs) 
\cite{Dudson,Huysmans}. A global magnetohydrodynamic (MHD) code {\tt NIMROD} 
\cite{Glasser} has also been applied to the modelling of ELMs \cite{Pankin}, in
addition to a range of other MHD instabilities in several different toroidal configurations
\cite{Sovinec}. In order to model turbulent transport on confinement and resistive diffusion 
timescales in an electromagnetic global code, it is necessary to include two-fluid effects, and it 
is also desirable to co-evolve the equilibrium.   

In this paper we describe \centori\/ (Culham Emulator of Numerical TORI), a new toroidal 
two-fluid, electromagnetic turbulence simulation code that meets these requirements. 
It can be used to describe the co-evolution of turbulence, MHD instabilities and equilibrium in
tokamak plasmas with arbitrary aspect ratio and high plasma beta (ratio of 
plasma pressure to magnetic field energy density). It is designed for the specific purpose 
of simulating global two-fluid electromagnetic tokamak plasma turbulence on
confinement timescales, in realistic geometries and in conditions such as
those found in the present-day machines MAST \cite{Meyer2009} and JET
\cite{Romanelli2009}, and in the forthcoming international fusion experiment
ITER \cite{Holtkamp2007}. Turbulent modes in tokamak plasmas are typically drift waves,
which are predominantly electrostatic waves driven by temperature or density gradients. 
An important example is the ion temperature gradient mode, which has wavelengths 
perpendicular to the magnetic field of the order of the ion Larmor radius $\rho_i$ \cite{Horton}. 
Many tokamak turbulence codes, such as the electrostatic fluid codes mentioned above
and also gyro-kinetic codes such as {\tt Kinezero} \cite{Romanelli}, are designed
specifically for the modelling of drift waves in a fixed, prescribed plasma equilibrium. 
\centori\/, on the other hand, is designed to study the interaction between drift waves and 
MHD instabilities, which generally occur at longer wavelengths, ranging up to the
system size, in a co-evolving equilibrium. However fluid codes such as \centori\/ cannot be 
used to model explicitly instabilities that occur on the smallest tokamak-relevant spatial scales, 
in particular length scales below the ion Larmor radius. Phenomena on the scale of the electron 
skin depth $\delta_e$ are specifically excluded from the model used in \centori\/, since electron 
inertia is neglected (in any event $\delta_e < \rho_i$ unless the plasma beta is less than the 
electron to ion mass ratio, which is not normally the case in the core region of tokamak plasmas). 
The drift waves described by gyro-kinetic theory have frequencies of the order of $\rho^*\Omega$  
where $\rho^*$ is particle Larmor radius normalised to the equilibrium gradient scale length and 
$\Omega$ is the corresponding cyclotron frequency \cite{Sugama}. Two-fluid theory, on the
other hand, can accommodate MHD modes such as global Alfv\'en eigenmodes 
\cite{McClements2002}, which, in low beta plasmas, have frequencies higher than 
those of ion drift waves. \centori\/ can be used to study processes occurring on timescales
ranging from the reciprocal Alfv\'en frequency to the energy confinement time. 

The physics model implemented in \centori\/ is very similar to
that used in {\tt CUTIE}, a global two-fluid electromagnetic turbulence code which was 
based on periodic cylinder geometry and was restricted to large aspect ratio 
plasmas with circular poloidal cross-section \cite{Thyagaraja2000}. Despite these 
restrictions, {\tt CUTIE} has been used for a number of successful applications. For 
example, it was recently shown to reproduce experimentally-observed transitions to a 
high confinement mode of plasma operation via the control of particle fuelling in the 
COMPASS-D tokamak \cite{Thyagaraja2010}.

This paper is organised as follows. In Section~\ref{sec:coordinates} we describe
the relationship between laboratory coordinates and plasma coordinates,
in which the fluid and Maxwell equations are evolved in \centori. The
form in which these equations are solved is discussed in
Sections~\ref{sec:quantities}--\ref{sec:normalised_physics_eqns}, while
initial and boundary conditions are discussed in
Section~\ref{sec:boundary_conditions}. Sections~\ref{sec:evolution}
and~\ref{sec:globals} are concerned respectively with the distinction made in
the code between mean and fluctuating quantities, and global quantities
evolved by it, such as plasma beta. In Section~\ref{sec:equilibrium_solver} we
describe \grass, a subroutine of \centori\/ which co-evolves the plasma
equilibrium using a novel pseudo-transient method. 
Operational and technical aspects of the \centori\ package and
the code structure are discussed in Section~\ref{sec:code_outline},
while in Section~\ref{sec:code_execution} we present some illustrative examples
of output from a simulation of a large aspect ratio tokamak plasma.

\section{Coordinate system}
\label{sec:coordinates}

Before describing the physical quantities and their evolution equations, it
is useful to provide a full description of the coordinate
systems used in \centori.

\subsection{Laboratory coordinates}

\centori\ is used to model a toroidal plasma held in place by magnetic fields
produced by external coils and by the plasma itself. A natural coordinate
system to use for the laboratory frame is the right-handed cylindrical system
$(R,Z,\zeta)$, where $R$ is major radius (distance from the machine's vertical
axis of symmetry), $Z$ is vertical distance (parallel to the symmetry axis),
and $\zeta$ is toroidal angle (azimuthal angle around the symmetry axis). We
note that
\begin{equation}
\grad\zeta =  -\nabla\phi = \frac{1}{R} \, \ezeta,
\label{eqn:gradzeta}
\end{equation}
where $\phi$ is azimuthal angle in the right-handed cylindrical
system $(R,\phi,Z)$ and $\ezeta$ is the unit vector in the $\zeta$ direction.

\subsection{Plasma coordinates}

The total magnetic field in the system comprises the vacuum field, produced
solely by currents flowing in conductors surrounding the plasma, plus the
field generated by the currents in the plasma itself. We use the total
equilibrium magnetic field to define the plasma coordinate system. The equilibrium
poloidal flux function $\psi(R,Z)$ defines the equilibrium poloidal magnetic field. 
The quantity $\psi(R,Z)$ can evolve in a \centori\ simulation, but only on a much longer 
timescale than the turbulence. It is
the magnetic flux per unit toroidal angle passing through the horizontal
circle of radius $R$ centred at $(R=0,Z)$; it is independent of $\zeta$.  When
plotted in the poloidal $(R,Z)$ plane the lines of constant $\psi$ in the
vicinity of the plasma form nested, closed contours (flux surfaces). The
minimum value of $\psi$ within these closed surfaces lies near the centre of
the plasma, and defines the location of the magnetic axis, along the circle
$(R_0,Z_0,\zeta)$.

\begin{figure}[ht]
\begin{center}
\epsfig{file=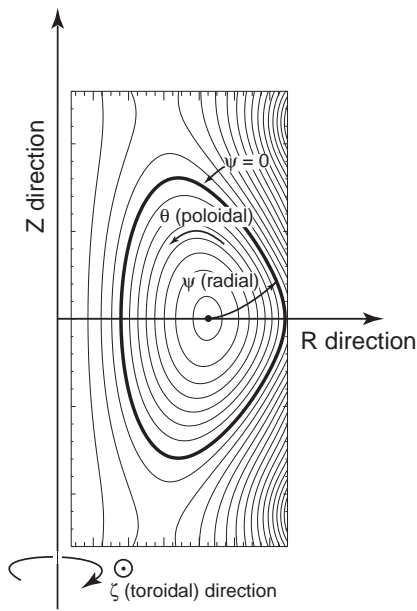,height=8cm,angle=0}
\parbox{14cm}{ \caption{\label{fig:psirz} Typical plot of $\psi$ contours over
    the $(R,Z)$ grid, showing laboratory coordinates $(R,Z,\zeta$) and plasma
    coordinates $(\psi,\theta,\zeta)$ employed in \centori. This plot was
    obtained using the \grass\/ equilibrium solver (Section~\ref{sec:grass}).}
}
\end{center}
\end{figure}

In a real machine the edge location of the plasma is determined by either a
physical limiter or the design of the magnetic geometry. Because only the
gradients of $\psi$ have physical meaning we may, for convenience, adjust
$\psi$ so that the known location of the edge of the plasma is defined to lie
on the $\psi=0$ contour. Figure~\ref{fig:psirz} shows a typical set of $\psi$
contours in the poloidal plane.
This plot is effectively the starting point for the calculations performed
using \centori.  The flux contours $\psi(R,Z)$ are determined \textit{a
  priori}\/ either by an external program or the equilibrium solver in the code, 
which is described in Section~\ref{sec:equilibrium_solver}; they are co-evolved in 
time with the turbulence.

Our aim is to evolve a set of plasma quantities which are stored in arrays at a
convenient set of computational grid points in a right-handed but in
general non-orthogonal dimensionless plasma coordinate system $(\rho,\theta,\zeta)$. Here 
$\rho$ is a radial coordinate, with $\nabla\rho$ directed from the magnetic 
axis to the plasma edge, and $\theta$ denotes an angle in the $(R,Z)$ plane. 

\subsection{Radial coordinate}

The radial coordinate $\rho$ is a normalised measure of $\psi$, the
normalising factor being the absolute value of the poloidal flux at the
magnetic axis, $\psi_0$. Thus, from the magnetic axis at $(R_0,Z_0)$ to the
edge of the plasma we have $-\psi_0 \leq \psi \leq 0$ and $0 \leq \rho \leq 1$
with $\rho$ defined in terms of $\psi$ by
\begin{equation}
\rho \equiv  1 + \psi/\psi_0 \; .
\label{eqn:rho}
\end{equation}
The radial grid points are equally spaced in $\rho$. In the cylindrical limit 
$\rho$ varies approximately as $r^2$ where $r$ is distance from the magnetic axis. The 
$\rho$ contours are thus relatively far apart near the magnetic axis, as
shown in Fig.~\ref{fig:psirz}. Because the magnetic axis is a
coordinate singularity (all $\theta$ points at $\rho=0$ and a given $\zeta$
coincide), we have chosen to locate the innermost $\rho$ grid points on a
contour that is slightly displaced from the axis itself.

The gradient $\grad\rho$ in the laboratory frame is
determined from the $\psi(R,Z)$ grid by fitting two-dimensional
Chebyshev polynomials \cite{Abramowitz} to the known $\psi$ values at the grid
points, and taking their derivatives in the $R$ and $Z$ directions. It follows from 
Eq.~(\ref{eqn:rho}) that
\begin{equation}
\grad\rho \equiv \frac{1}{\psi_0} \grad\psi = \frac{1}{\psi_0} \frac{\dpsi}{\dd R} \er + 
\frac{1}{\psi_0} \frac{\dpsi}{\dd Z} \ez \label{eqn:gradrho2},
\end{equation}
where $\er$ and $\ez$ denote unit vectors in the $R$ and $Z$ directions.

\subsection{Relationship between equilibrium magnetic field and plasma 
  coordinates}

The equilibrium poloidal magnetic field is given by
\begin{equation}
\B_p \equiv \grad\zeta \X \grad\psi = \psi_0 \, (\grad\zeta \X \grad\rho).
\label{eqn:bp}
\end{equation}
Thus $B_p = \psi_0\vert\nabla\rho\vert/R$. The toroidal equilibrium magnetic
field is given by
\begin{equation}
\B_t \equiv F \, \grad\zeta,
\label{eqn:F}
\end{equation}
where the scalar quantity $F$ is taken to be a flux function, i.e.\ it depends
only on the radial coordinate $\rho$. This is generally a good approximation
under typical tokamak conditions \cite{McClements2011}. Thus the total
equilibrium magnetic field is
\begin{equation}
\Beq = \psi_0\,(\grad\zeta \X \grad\rho) + F\,\grad\zeta .
\label{eqn:beq}
\end{equation}
We define a vector potential $\A$ in the usual way as a vector field whose
curl is equal to the magnetic field. We can write the equilibrium vector
potential $\Aeq$ in covariant form as follows:
\begin{equation}
\Aeq = A_{\mbox{eq}\,\rho} \grad\rho + A_{\mbox{eq}\,\theta} \grad\theta
+ A_{\mbox{eq}\,\zeta} \grad\zeta .
\end{equation}
For convenience we choose a gauge such that the radial component of $\Aeq$
vanishes, i.e.
\begin{equation}
A_{\mbox{eq}\,\rho} = 0.
\label{eqn:arho}
\end{equation} 
In terms of the remaining components of $\Aeq$, the equilibrium 
magnetic field becomes
\begin{equation}
\Beq = \curl\Aeq = \grad A_{\mbox{eq}\,\theta} \X \grad\theta
+ \grad A_{\mbox{eq}\,\zeta} \X \grad\zeta .
\label{eqn:tmp04}
\end{equation}
Matching the poloidal components of Eqs.~(\ref{eqn:beq}) and~(\ref{eqn:tmp04})
we find that we can set
\begin{equation}
A_{\mbox{eq}\,\zeta} = -\psi.
\label{eqn:azeta}
\end{equation} 
Matching the toroidal components of Eqs.~(\ref{eqn:beq}) and~(\ref{eqn:tmp04})
we obtain
\begin{equation}
F \grad\zeta = \grad A_{\mbox{eq}\,\theta} \X \grad\theta
 = \frac{\dd A_{\mbox{eq}\,\theta}}{\drho} \grad\rho \X \grad\theta ,
\end{equation}
and the scalar product of this with $\grad\zeta$ yields
\begin{equation}
F \grad\zeta\cdot\grad\zeta = {F\over R^2} = 
\frac{\dd A_{\mbox{eq}\,\theta}}{\drho} \grad\zeta\cdot(\grad\rho \X
\grad\theta) = \frac{\dd A_{\mbox{eq}\,\theta}}{\drho}\jac,
\end{equation}
where $\jac \equiv \grad\zeta\cdot(\grad\rho\times\grad\theta)$ is the
Jacobian relating laboratory and plasma coordinates (see following
subsection). The covariant poloidal component of $\Aeq$ is thus given by
\begin{equation}
A_{\mbox{eq}\,\theta} = \int \frac{F}{\jac R^2} d\rho.
\label{eqn:atheta}
\end{equation}
Eqs.~(\ref{eqn:arho}),~(\ref{eqn:azeta}) and~(\ref{eqn:atheta}) define the
equilibrium vector potential $\Aeq$ in covariant form; the equilibrium
magnetic field $\Beq$ may be calculated by taking its curl.

The set of space variables $(\rho,\theta,\zeta)$ constitutes a
quasi-orthogonal coordinate system in which $\grad\rho\cdot\grad\zeta =
\grad\theta\cdot\grad\zeta = 0$, but in general $\grad\rho\cdot\grad\theta
\not= 0$. Taking scalar products of $\Beq$ with the coordinate 
gradients we obtain
\begin{equation}
\Beq\cdot\grad\rho = 0, \;\;\;\;\;
\Beq\cdot\grad\theta = \psi_0\,(\grad\zeta \X \grad\rho)\cdot\grad\theta = \psi_0\,\jac,
\;\;\;\;\;
\Beq\cdot\grad\zeta = F/R^2.
\end{equation}
These three equations give the contravariant components of $\Beq$ directly,
thereby eliminating the need to perform a curl operation (see
Section~\ref{sec:coordproperties}).

\subsection{Poloidal coordinate}

\begin{figure}[ht]
\begin{center}
\epsfig{file=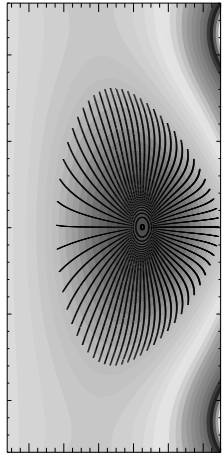,height=6cm,angle=0}
\parbox{14cm}{
\caption{\label{fig:coords} Typical set of $(\rho,\theta)$ grid points, superimposed on the
original $\psi(R,Z)$ grid. In this case there are 129 radial grid points and
65 poloidal grid points.}
}
\end{center}
\end{figure}

The poloidal angle $\theta$ varies from $0$ to $2\pi$ in the $(R,Z)$ plane. By
convention, points at $\theta = 0$ lie along the line defined by $(R \geq R_0,
Z = Z_0)$, and $\theta$ increases in the anticlockwise direction as shown in
Fig.~\ref{fig:psirz}. Denoting by $l$ the arc length in the poloidal plane along
a given $\rho$ contour, we can write
\begin{equation}
\Beq\cdot\grad\theta = B_p \frac{\dtheta}{\dd l} 
 = \psi_0 \, \frac{|\grad\rho|}{R} \frac{\dtheta}{\dd l} 
= \psi_0 \, \jac .
\label{eqn:dtheta}
\end{equation}
To determine the distribution of $\theta$ grid points along the contour in the
$(R,Z)$ plane we introduce a parameter $\tau$ and solve the following pair of
Hamiltonian equations \cite{McClements2011}:
\begin{equation}
\frac{dR}{d\tau} = -\frac{\drho}{\dd Z}
\; ; \;\;\;
\frac{dZ}{d\tau} = \frac{\drho}{\dd R},
\end{equation}
with $(R,Z)$ being stored at intermediate points as the solution proceeds. The
gradients in $\rho$ are calculated using Chebyshev polynomials, as described
above, and a convergence loop ensures that the contour is followed with
sufficient accuracy. The arc length $l$ is given in terms of $\tau$ by
\begin{equation}
\frac{dl}{d\tau} =
\sqrt{ \left( \frac{dR}{d\tau} \right)^2 +
       \left( \frac{dZ}{d\tau} \right)^2 } = 
\sqrt{ \left( \frac{\drho}{\dd R} \right)^2 +
       \left( \frac{\drho}{\dd Z} \right)^2 } = |\grad\rho|.
\end{equation}
We choose $\jac$ to be a flux function, i.e.\ $\jac =
\jac(\rho)$. This enables $\theta$ points on a given flux contour to be
determined by integrating the expression
\begin{equation}
d\theta = R\,\jac\,\frac{dl}{|\grad\rho|} = R\,\jac\,d\tau,
\end{equation}
where $\jac$ is obtained by imposing a 2$\pi$ periodicity on $\theta$:
\begin{equation}
\jac(\rho) = \frac{2\pi}{\oint R\,d\tau}.
\label{eqn:jacobian2}
\end{equation}
It is straightforward to interpolate the stored $(R,Z)$ values to determine the
locations of equally-spaced $\theta$ points along the $\rho$
contour. Figure~\ref{fig:coords} shows an example of a
$(\rho,\theta)$ grid.

The process described above can be used to map out the locations $R(\rho,\theta)$,
$Z(\rho,\theta)$ along the $\rho$ contours. The partial derivatives $\dd R/\drho$, 
$\dd R/\dtheta$, $\dd Z/\drho$ and $\dd Z/\dtheta$ are found by fitting Chebyshev 
polynomials to $R$ and $Z$ along the $\rho$ direction and Fourier series in
the $\theta$ direction. These provide the contravariant basis vectors of the plasma
coordinate system:
\begin{equation}
\mathbf{b}_\rho \equiv \jacstar\,(\grad\theta \X \grad\zeta)
= \frac{\dd R}{\drho}\,\er + \frac{\dd Z}{\drho}\,\ez,
\end{equation}
\begin{equation}
\mathbf{b}_\theta \equiv \jacstar\,(\grad\zeta \X \grad\rho)
= \frac{\dd R}{\dtheta}\,\er + \frac{\dd Z}{\dtheta}\,\ez,
\end{equation}
\begin{equation}
\mathbf{b}_\zeta \equiv \jacstar\,(\grad\rho \X \grad\theta)
= R\,\ezeta,
\end{equation}
where $\jacstar = 1/\jac$ and $\mathbf{b}_\zeta$ follows directly from 
Eq.~(\ref{eqn:gradzeta}); $\grad\zeta$ is the covariant $\zeta$ basis vector
and hence is reciprocal to $\mathbf{b}_\zeta$. We may then calculate $\jacstar$ 
(and therefore $\jac$) using
\begin{equation}
\jacstar = \mathbf{b}_\theta\cdot(\mathbf{b}_\zeta \X \mathbf{b}_\rho),
\end{equation}
and the covariant basis vectors are given by
\begin{equation}
\mathbf{b}^\rho \equiv
\grad\rho = \jac\,(\mathbf{b}_\theta \X \mathbf{b}_\zeta), \;\;\;\;\;
\mathbf{b}^\theta \equiv
\grad\theta = \jac\,(\mathbf{b}_\zeta \X \mathbf{b}_\rho), \;\;\;\;\; 
\mathbf{b}^\zeta \equiv
\grad\zeta = \frac{1}{R} \ezeta.
\end{equation}
It is straightforward to evaluate these vector products since the covariant
and contravariant basis vectors are all stored with components in the
laboratory frame (although they are evaluated at specified
points in $(\rho,\theta)$ space, their components relative to the
basis $(\er,\ez,\ezeta)$ are known).

We adopt this particular algorithm to obtain the gradients and the Jacobian in
order to maximise accuracy and smoothness in the results through the use of
the Chebyshev/Fourier fitting method, and it also guarantees that the
covariant and contravariant basis vectors are reciprocal.

\subsection{Vector operations in plasma coordinate system}
\label{sec:coordproperties}

This section provides expressions for scalar and vector products together with differential
operators in the plasma coordinate system. In what follows $\A$ and $\B$ are 
arbitrary vector functions, while $f$ is an arbitrary scalar function.
The vector $\A$ has covariant representation
\begin{equation}
\A = A_\rho\,\bcovrho + A_\theta\,\bcovtheta + A_\zeta\,\bcovzeta,
\end{equation}
where $A_i = {\bf A}\cdot{\bf b}_i$. The corresponding contravariant representation is
\begin{equation}
\A = A^\rho\,\bconrho + A^\theta\,\bcontheta + A^\zeta\,\bconzeta,
\end{equation}
where $A^i = {\bf A}\cdot{\bf b}^i$. The scalar product of $\A$ and $\B$ is then 
$\A\cdot\B = A_iB^i = A^iB_i$, where a repeated index implies summation, while the vector product 
is given by
\begin{eqnarray}
\A \X \B & = & \;\, \jac \left\{ \rule{0mm}{4mm}
(A_\theta B_\zeta - A_\zeta B_\theta) \bconrho + 
(A_\zeta B_\rho - A_\rho B_\zeta) \bcontheta + 
(A_\rho B_\theta - A_\theta B_\rho) \bconzeta \right\} \nonumber \\
& = & \jacstar \left\{ \rule{0mm}{4mm}
(A^\theta B^\zeta - A^\zeta B^\theta) \bcovrho + 
(A^\zeta B^\rho - A^\rho B^\zeta) \bcovtheta + 
(A^\rho B^\theta - A^\theta B^\rho) \bcovzeta \right\}.
\end{eqnarray}
The gradient operator, which is defined in the usual way, produces a covariant vector.
The divergence of $\A$ is evaluated using its contravariant components:
\begin{equation}
\nabla\cdot\A = \jac\, \left\{
\frac{\dd}{\drho} \left( \frac{A^\rho}{\jac} \right) + 
\frac{\dd}{\dtheta} \left( \frac{A^\theta}{\jac} \right) + 
\frac{\dd}{\dzeta} \left( \frac{A^\zeta}{\jac} \right) \right\},
\end{equation}
while the curl is obtained using its covariant components, and the result is a
contravariant vector:
\begin{equation}
\nabla \X \A = \jac\, \left\{
\left( \frac{\dd A_\zeta}{\dtheta} - \frac{\dd A_\theta}{\dzeta} \right)
\bconrho +
\left( \frac{\dd A_\rho}{\dzeta} - \frac{\dd A_\zeta}{\drho} \right)
\bcontheta +
\left( \frac{\dd A_\theta}{\drho} - \frac{\dd A_\rho}{\dtheta} \right)
\bconzeta \right\}.
\end{equation}
The choice of $\jac$ as a flux function considerably simplifies and speeds up many
calculations.  

\subsection{Physical coordinates}
\label{sec:physical}

It is convenient to perform the vector operations discussed above using the
covariant and contravariant representations. However, these do not have
directions, dimensions or units that are intuitive as far as the physics is
concerned. We therefore define a third set of components for the vector
quantities in \centori, which we refer to as their \textit{physical
  representation}: ``normal'', denoting the direction normal to the flux surface;
``tangential'', parallel to the flux surface
in the $(R,Z)$ plane; and ``toroidal'', around the 
machine axis. The physical components are orthogonal:
\begin{equation}
A_\mathbf{normal} = \A\cdot \frac{\grad\rho}{|\grad\rho|} =
\frac{A^\rho}{|\grad\rho|}, 
\end{equation}
\begin{equation}
A_\mathbf{tangential} =
\A\cdot \frac{(\grad\zeta \X \grad\rho)}{|\grad\zeta \X \grad\rho|} =
\A\cdot \frac{(\grad\zeta \X \grad\rho)}{|\grad\rho|/R} =
A_\theta\, \frac{\jac\,R}{|\grad\rho|},
\end{equation}
\begin{equation}
A_\mathbf{toroidal} = \A\cdot \frac{\grad\zeta}{|\grad\zeta|} =
\frac{A^\zeta}{|\grad\zeta|}.
\end{equation}

\subsection{Flux surface-averaged quantities}

It is necessary to compute flux surface-averaged quantities in \centori\/ since these
affect the evolving equilibrium. The flux surface average of a scalar quantity 
$f(\rho,\theta,\zeta)$ is given by
\begin{equation}
\langle f \rangle_\rho = \frac{\int\!\!\!\int f \, d\theta \, d\zeta/\jac}
{\int\!\!\!\int \, d\theta \, d\zeta/\jac} = {1\over 4\pi^2}\int\!\!\!\int f \, d\theta \, d\zeta, 
\label{eqn:fluxsurfaceaverage}
\end{equation}
where $f$ is evaluated at fixed $\rho$ and we have used the fact that $\jac$ is 
defined to be a flux function.  

\section{Physics quantities}
\label{sec:quantities}

The primary quantities evolved by \centori\/ are as follows:
$\vi$, ion velocity; $\A$, vector potential; $n_i$, ion number density ($ = n_e$, electron 
number density, via quasi-neutrality); $T_i$, ion temperature; and $T_e$, electron temperature.
In addition, a number of auxiliary quantities can be advanced in time once the primary 
quantities have been updated. These are: $\ve$, electron velocity; $\Phi$, electric potential; 
$\B$, magnetic field; $\J$, current density; $\E$, electric field; $p_i = n_e \, T_i$, 
ion pressure; and $p_e = n_e \, T_e$, electron pressure.
              
These variables are normalised as follows:
\begin{equation}
\vistar = \frac{\vi}{\va}, \mbox{\hspace{0.8cm}}
\vestar = \frac{\ve}{\va}, \mbox{\hspace{0.8cm}} 
\Astar = \frac{\A}{B_0}, \mbox{\hspace{0.8cm}}
\Bstar = \frac{\B}{B_0}, \mbox{\hspace{0.8cm}}
\label{eqn:varnorm1}
\end{equation}
\begin{equation}
\nestar = \frac{n_e}{\nave}, \mbox{\hspace{0.8cm}} 
\tistar = \frac{T_i}{T_{i0}}, \mbox{\hspace{0.8cm}}
\testar = \frac{T_e}{T_{e0}}, \mbox{\hspace{0.8cm}}
\pistar = \nestar \, \tistar = \frac{p_i}{p_{i0}},  \mbox{\hspace{0.8cm}} 
\pestar = \nestar \, \testar = \frac{p_e}{p_{e0}},
\label{eqn:varnorm2}
\end{equation}
where $\va = B_0/\sqrt{4\pi \rho_m} \simeq B_0/\sqrt{4\pi\,m_i\,\nave}$ is a
typical Alfv\'en speed, $\rho_m = m_i\,n_e$ is the ion mass density, $B_0$ is
the vacuum toroidal field at the magnetic axis, $\nave$ is the volume-averaged
electron number density, $T_{i0}$ is the initial ion temperature at the
magnetic axis, $T_{e0}$ is the initial electron temperature at the magnetic
axis, $p_{i0} = \nave\,T_{i0}$ is a nominal ion pressure,
$p_{e0} = \nave\,T_{e0}$ is a nominal electron pressure, and
$m_i$ is the ion mass. The quantities $B_0$, $T_{i0}$ and $T_{e0}$ are given
nominal values by the user; $\nave$ is calculated as the plasma evolution
progresses, so $\va$, $p_{i0}$ and $p_{e0}$ vary with time. It should be noted that the
actual density and temperature values on axis are not constrained
to be their initial arbitrary values, but vary as the profiles evolve. The normalised quantities 
listed above are all dimensionless except for $\Astar$ which has the dimensions of length.

\section{Two-fluid equations}
\label{sec:twofluidequations}

\subsection{Momentum equations}

The ion momentum balance equation can be written in the form
\begin{equation}
\rho_m \left( \frac{\dd \vi}{\dt} + \W \X \vi \right) = 
-\grad p_i - \frac{\rho_m}{2} \grad\vi^2 + en_e{\bf E} + \frac{en_e}{c} (\vi \X \B)
- en_e\eta \J - \rho_m\,\chi_v (\curl \W) + \Sv.
\label{eqn:momentum}
\end{equation}
where $\W = \curl \vi$ is vorticity, $\eta$ is resistivity (assumed to be a scalar 
function of space and time), $\chi_v$ is velocity diffusivity (see 
Section~\ref{sec:momentum_source}), $\Sv$ is external force density 
(see Section~\ref{sec:momentum_source}), $e$ is proton charge and $c$ is the speed of light
(we use Gaussian cgs units throughout this paper, although output from the code is in SI 
units, to facilitate comparison with experimental results). 
In the electron momentum balance equation we neglect inertial terms, momentum sources and 
viscosity:
\begin{equation}
{\bf 0} = -\nabla p_e-en_e{\bf E}-{en_e\over c}({\bf v_e}\times{\bf B})+en_e\eta{\bf J}.
\label{eqn:ohms_law}
\end{equation}
This is equivalent to Ohm's law in the limit of vanishing electron mass.
 
\subsection{Energy equations}

The transfer of energy is described by the two equations
\begin{equation}
\frac{3}{2}n_e\left(\frac{\dd}{\dt} + \vi\cdot\nabla\right)T_i + p_i \nabla\cdot\vi = 
-\nabla\cdot\q_i + S_i,
\end{equation}
\begin{equation}
\frac{3}{2} n_e\left(\frac{\dd}{\dt} + \ve\cdot\nabla\right)T_e + p_e \nabla\cdot\ve = 
-\nabla\cdot\q_e + S_e,
\label{eqn:energy}
\end{equation}
where $\q_{i,e}$ are the ion and electron heat fluxes (see 
Section~\ref{sec:normalised_energy_eqns}) and $S_{i,e}$ are additional ion and 
electron heating sources.

\subsection{Mass continuity equation}

The mass continuity equation used in \centori\/ is
\begin{equation}
\frac{\dd \rho_m}{\dt} + \grad\cdot(\rho_m \vi) = S_n -
m_i\,\nave\,\va\,\nabla\cdot\Gamma_W^* + \delta_n -
\nuparai (\rho_m - \langle \rho_m \rangle),
\label{eqn:continuity_eqn} 
\end{equation}
where $S_n$ is the particle source rate (see Section~\ref{sec:particle_source}), 
$\nabla\cdot\Gamma_W^*$ is a term representing the effect of the Ware pinch \cite{Ware} 
(see Section~\ref{sec:normalised_energy_eqns}), $\delta_n$ is a diffusion term given by
\begin{equation}
\delta_n = \frac{2}{3} R_0\,\jac \left\{
\left( (\chi_{ne}+\chi_{RR}) \frac{\dd^2 \rho_m}{\drho^2} \right)
+ \frac{\langle B_p \rangle^2}{B_p^2} \left( \chi_{ne} \frac{\dd^2
\rho_m}{\dtheta^2} \right)
\right\},
\label{eqn:n_diffusion}
\end{equation}
with $\chi_{ne}$ and $\chi_{RR}$ respectively the particle and Rechester-Rosenbluth
diffusivities (see Sections~\ref{sec:normalised_momentum_eqn} and \ref{sec:particle_diffusion}).
Finally in Eq.~(\ref{eqn:continuity_eqn}), $\nuparai$ is the parallel ion thermal relaxation 
rate (see Section~\ref{sec:normalised_energy_eqns}). It is not necessary to solve an electron 
continuity equation since the plasma is assumed to be quasi-neutral and the current is assumed
to be divergence-free.   

\subsection{Maxwell's equations}

The vanishing of the divergence of {\bf B} is guaranteed in \centori\/
through the use of the potential representation $\B = \curl \A$ and the induction equation is 
solved for {\bf A} rather than {\bf B}:
\begin{equation}
\frac{1}{c} \frac{\dd \A}{\dt} = -\E - \grad\Phi.
\label{eqn:faradays_law}
\end{equation}
Current densities ${\bf J}$ are computed using the pre-Maxwell form of Amp\`ere's law:
\begin{equation}
\J = \frac{c}{4\pi} \curl \B .
\label{eqn:amperes_law}
\end{equation}

\section{Normalised physics equations and their solution}
\label{sec:normalised_physics_eqns}

\centori\/ is used to evolve the normalised quantities defined by Eqs.~
(\ref{eqn:varnorm1}) and (\ref{eqn:varnorm2}) rather than the absolute values of
velocity, magnetic field, and so on. In this section we explain how the
physics equations themselves are normalised. Unless otherwise stated, all of the 
normalised equations have the dimensions of reciprocal length. The equations are solved 
by using finite differences to approximate all of the derivatives; the solution method 
is thus entirely non-spectral. A key advantage of this approach is that parallelisation 
of the code is then relatively straightforward, and yields good scalability results 
(see Section~\ref{sec:code_outline}). On the other hand the finite-element method, 
used, for example, in {\tt NIMROD} \cite{Glasser}, is particularly well-suited to modelling 
the edge regions of plasmas with strongly-shaped poloidal cross-sections.

\subsection{Normalised ion momentum equation}
\label{sec:normalised_momentum_eqn}

All three physical components of $\vistar$ are evolved, with subscript ``1''
labelling the normal direction, subscript ``2'' the tangential direction, and
subscript ``3'' the toroidal direction. We define a normalised vorticity $\Wstar$:
\begin{equation}
\Wstar \equiv\frac{\W}{\va} = (\curl \vistar).
\label{eqn:phistar}
\end{equation}
It should be noted that $\Wstar$ has the dimensions of reciprocal length.
Using also the normalizations introduced previously, dividing by $\va^2m_i\nave\nestar$,
defining the following quantities
$$ \beta_{i0} \equiv \frac{4\pi\,p_{i0}}{B_0^2}
  = \frac{p_{i0}}{\va^2\,m_i\,\nave}, \;\;\;\;\;\; \beta_{e0} \equiv \frac{4\pi\,p_{e0}}{B_0^2} 
= \frac{T_{e0}}{\va^2\,m_i}, \;\;\;\;\;\; D_v \equiv \frac{\chi_v}{\va}, \;\;\;\;\;\; \Svstar 
\equiv \frac{\Sv}{\va^2\,m_i\,\nave} = \frac{4\pi\,\Sv}{B_0^2}, $$
and introducing an additional term related to the Rechester-Rosenbluth diffusivity $D_{RR}$
\cite{Rechester} [see Eq. (\ref{eqn:chirr})], we find that the ion momentum equation can be 
written in the form
\begin{eqnarray}
\frac{1}{\va} \frac{\dd \vistar}{\dt} & = &
- \left[ \Wstar + \frac{\omega_{ci}}{\va} \Bstar \right] \X \vistar
- \beta_{i0} \frac{\grad\pistar}{\nestar}
- \beta_{e0} \grad\Phistar
- \frac{1}{2} \grad \vistar^2
+ \frac{\Svstar}{\nestar} \nonumber \\
&& + \frac{\omega_{ci}}{\va} \Beqstar \X \left( D_{RR}
    \frac{\grad \langle \nestar \rangle} {\langle \nestar \rangle} \right)
- D_v \, (\curl \Wstar)
- {\omega_{ci}\over \va}\left[{1\over \va}\frac{\dd \Astar}{\dt}
+ \eta^* \Jstar\right],
\label{eqn:normalised_momentum_eqn}
\end{eqnarray}
where $\Jstar \equiv 4\pi\J/(cB_0)$, $\eta^* \equiv c^2\eta/(4\pi\va)$, and 
$\omega_{ci} = eB_0/(m_ic)$. To reduce problems arising from short wavelength modes in the radial
direction, the momentum equation is supplemented by artificial damping terms:
\begin{equation}
\frac{1}{\va} \frac{\dd}{\dt} \left(v_{i,\,\mbox{normal}}^*\right) = 
\ldots - \delta_v \, v_{i,\,\mbox{normal}}^*,
\end{equation}
where $\delta_v = 0.5\nuparai/\va$, $\nuparai$ being the parallel ion thermal 
relaxation rate [Eq.~(\ref{eqn:nuparai})]. A similar damping term is applied
in the tangential direction. The dimensionless damping rate used in the 
code is $\delta_v^* = \va\delt\delta_v$.

\subsubsection{Evolution of normalised momentum equation}
\label{sec:vi_evolution}

In the current version of \centori\ we neglect the $(1/\va)\partial{\bf A}^*/\partial t$ and
${\bf J}^*$ terms on the right hand side of Eq.~(\ref{eqn:normalised_momentum_eqn}). In 
tokamak plasmas there is generally a large separation between drift and Alfv\'en 
timescales, with the consequence that turbulent fluctuations are predominantly electrostatic in 
nature, and the inductive part of the electric field term plays only a minor role in the ion 
momentum equation. In Section~\ref{sec:turbsim} we will illustrate this point using results from a 
\centori\ simulation. The ${\bf J}^*$ term in Eq.~(\ref{eqn:normalised_momentum_eqn}) 
is small by virtue of the fact that tokamak plasmas tend to have very high Lundquist numbers, i.e.
are close to being perfectly conducting.   

In Eq. (\ref{eqn:normalised_momentum_eqn}) it is not straightforward to convert $\delta \equiv 
-D_v (\curl \Wstar)$ into finite differences, due to the non-orthogonal nature of the 
coordinate system. We approximate it by the expression
\begin{equation}
\delta \simeq R_0\,\jac\,D_v \left(
\frac{\dd^2 \vistar}{\drho^2} + \frac{\langle B_p \rangle^2}{B_p^2}
\frac{\dd^2 \vistar}{\dtheta^2} \right).
\label{eqn:vi_diffusion}
\end{equation}
We are assuming here that the contribution of viscosity to the ion momentum equation can be well-approximated 
by a term proportional to $\nabla^2{\bf v}_i^*$. We are thus neglecting the $\nabla(\nabla\cdot{\bf v}_i^*)$ term in 
$\nabla\times{\bf W}^*$ (although the flows described by \centori\/ are in general compressible); it is not necessary to 
include this term in order to model the neoclassical and turbulent damping of flows \cite{Thyagaraja2005}. 
The factor containing $B_p$ is present to take account of the spacing of
adjacent points in the $\theta$ direction being proportional to the local
poloidal field [see Eq. (\ref{eqn:dtheta})]. Adopting the convention that subscripts 
$j$, $k$ and $l$ label array elements in the $\rho$, $\theta$ and $\zeta$ directions 
respectively, while superscripts $N$ label time, we approximate $\delta$ by the 
central difference expression 
\begin{eqnarray}
\delta & = & \frac{R_0\,\jac_{j,k}\,D_{v\,j,k,l}}{(\delrho)^2} \left(
\vistar_{j+1,k,l}^{N+1} + \vistar_{j-1,k,l}^{N+1} - 
2 \vistar_{j,k,l}^{N+1} \right) \nonumber \\
&& + \frac{R_0\,\jac_{j,k}\,D_{v\,j,k,l}}{(\deltheta)^2} \frac{\langle B_p 
\rangle^2}{B_{p\,j,k,l}^2} \left(
\vistar_{j,k+1,l}^{N} + \vistar_{j,k-1,l}^{N} - 2 \vistar_{j,k,l}^{N+1} \right).
\end{eqnarray}
We also introduce purely numerical diffusion terms, with coefficients
$\epsilon_{\rho}$, $\epsilon_{\theta}$ and $\epsilon_{\zeta}$, which are
designed to remove variations in $\vistar$ of similar length scale to the
separation of adjacent grid points. It is evident that the resultant
finite-difference equations are consistent with the governing partial
differential equations as the mesh sizes tend to infinity. These effectively
suppress spurious oscillations at wave numbers corresponding to inverse mesh
size. Unlike the turbulent diffusivities, they are non-zero even when the
turbulent fluctuation amplitudes go to zero for any fixed mesh size. Thus the
finite difference approximation to the momentum equation is of the form
\begin{eqnarray}
\vistar_{j,k,l}^{N+1} & = & \vistar_{j,k,l}^{N} + \frac{\epsilon_\rho}{2}
\left(\vistar_{j+1,k,l}^{N+1} + \vistar_{j-1,k,l}^{N+1} - 2
\vistar_{j,k,l}^{N+1}\right) \nonumber \\
&& + \frac{\epsilon_\theta}{2} \left(\vistar_{j,k+1,l}^{L} + \vistar_{j,k-1,l}^{L} - 2
\vistar_{j,k,l}^{N+1}\right)
+ \frac{\epsilon_\zeta}{2} \left(\vistar_{j,k,l+1}^{L} + \vistar_{j,k,l-1}^{L} - 2
\vistar_{j,k,l}^{N+1}\right).
+ \mbox{\ldots}
\label{eqn:numdiff}
\end{eqnarray}
Here, superscripts $L$ (``latest'') indicate the most up-to-date (most
time-advanced) values available. This is to avoid the use of ``new'' values at
adjacent $\theta$ and $\zeta$ points (i.e.\ at $k \pm 1$, $l \pm 1$); these
would appear as undesirable off-diagonal terms in the tridiagonal matrix
equation. Typically, $\vistar^L \equiv\vistar^{N+1}$ from the previous
iteration.  The numerical diffusion coefficients $\epsilon_\rho$,
$\epsilon_\theta$ and $\epsilon_\zeta$ have the following forms:
$$ \epsilon_\rho = \frac{\sqrt{\rho}}{N_\psi^2}, \;\;\;\;\;\; \epsilon_\theta = 
\frac{\sqrt{\rho}}{2\pi^2 N_\theta^2}, \;\;\;\;\;\; \epsilon_\zeta = \frac{1}{8\pi^2 N_\zeta^2}, $$
where $N_\psi$, $N_\theta$ and $N_\zeta$ are the numbers of grid
intervals in the respective directions.

Dropping the $k$ and $l$ subscripts on velocity components, the finite difference approximation to 
the momentum equation can be written in the block tridiagonal matrix equation form
\begin{equation}
\underline{\mathbf{A}}_j \vistar_{j-1}^{N+1} 
+ \underline{\mathbf{B}}_j \vistar_{j}^{N+1} 
+ \underline{\mathbf{C}}_j \vistar_{j+1}^{N+1} 
= \mathbf{R}_j, 
\label{eqn:block_tdma}
\end{equation}
where $\underline{\mathbf{A}}_j$, $\underline{\mathbf{B}}_j$, $\underline{\mathbf{C}}_j$
are $3 \X 3$ matrices and $\mathbf{R}_j$ is a vector that depends on the latest ($L$) values
of velocity components as well as those of the previous timestep. Equation (\ref{eqn:block_tdma}) 
is solved for the normalised ion velocity $\vistar$ at the new timestep using a standard 
predictor-corrector scheme, with $\vistar_{j}^{L}$ converging to $\vistar_{j}^{N+1}$. We 
then subtract the flux surface-averaged normal component of $\vistar$, so that only a 
fluctuating part remains.

\subsubsection{Momentum sources and transport}
\label{sec:momentum_source}

Presently only a toroidal momentum source is included in \centori; the profile is given by
\begin{equation}
S_{v,tor}^*(\rho) \equiv f_{mom} \, \frac{4\pi}{B_0^2} \,
\frac{P_{aux,i}(\rho)+P_{aux,e}(\rho)}{v_{th,i}(0)},
\end{equation}
where $P_{aux,i/e}$ is the external heating power per unit volume provided to
the ions/electrons, $v_{th,i}(0) = (2T_i(0)/m_i)^{1/2}$ is the ion thermal
speed at the magnetic axis and $f_{mom}$ is a user-defined multiplier which
matches the total momentum provided to the plasma with experiment.

Turbulent diffusivity terms are included in the full definition of the
normalised velocity diffusivity $D_v$ [see Eq. (\ref{eqn:normalised_momentum_eqn})]:
\begin{equation}
D_v(\rho,\theta,\zeta) = \frac{\chi_{v,\mbox{\scriptsize user}}}{\va}
\left( 1 + q \langle R \rangle^2
\sqrt{\frac{m_i}{m_e}} \left[ f_{JJ}\,\Jfstar^2 + \Wfstar^2
\right] \right)
+ \frac{\chi_{v,\mbox{\scriptsize classical}}}{\va},
\end{equation}
where $\chi_{v,\mbox{\scriptsize user}}$ is user-specified, $\Jfstar^2$ is a
normalised measure of entropy and $\Wfstar^2$ is a normalised enstrophy 
($\Jfstar$ and $\Wfstar$ being the fluctuating parts of the normalised current 
density and vorticity respectively). The parameter $f_{JJ}$ is a user-defined multiplier
between 0 and 1. The final term in $D_v$ is a Gyro-Bohm diffusivity:
\begin{equation}
\chi_{v,\mbox{\scriptsize classical}}(\rho) = f_{\chi c} \, \frac{v_{th,i}\,
\rho_i^2}{a},
\end{equation}
with $0 \leq f_{\chi c} \leq 1$ a user-defined multiplier and $\rho_i$ the ion
gyroradius.

\subsection{Evolution of normalised electron velocity}
\label{sec:ve_evolution}

The normalised electron velocity $\vestar$ is determined directly from
$\vistar$ and $\Jstar$ by noting that the net current density $\J$ is given by
\begin{equation}
\J = e\,n_e\,(\vi - \ve).
\label{eqn:ve}
\end{equation}
Hence
\begin{equation}
\vestar = \vistar - \frac{c\,B_0}{4\pi\,e\,\nave\,\nestar\,\va} \Jstar.
\label{eqn:vestar}
\end{equation}

\subsection{Evolution of electromagnetic quantities}

\subsubsection{Normalised Amp\`{e}re's Law}

It is evident from the definitions of $\Bstar$ and $\Jstar$ that Amp\`{e}re's
law has the normalised form
\begin{equation}
\Jstar = \curl \Bstar.
\label{eqn:jstar2}
\end{equation}
Note that $\Bstar$ is dimensionless while $\Jstar$ has the dimensions of reciprocal length.

\subsubsection{Normalised Faraday law and Ohm's law}

Ohm's law [Eq.~(\ref{eqn:ohms_law})] can be written in the form
\begin{equation}
\E = -\frac{\ve \X \B}{c} -\frac{\grad p_e}{e\,n_e}+\eta \J.
\label{eqn:ohms_law1}
\end{equation}
We divide the electric field into ideal and resistive parts by writing
$$ \E_{\rm ideal} = -\frac{\ve \X \B}{c} -\frac{\grad p_e}{e\,n_e}, $$
$$ \E_{\rm res} = \eta \J, $$
and write Faraday's law in the form
$$ \frac{1}{c} \frac{\dd \A}{\dt} = - \E_{\rm ideal} - \grad\Phi - \E_{\rm res} = - \left(
- \frac{\ve \X \B}{c}
- \frac{\grad p_e}{e\,n_e} \right)
- \grad\Phi - \eta \J. $$
We can thus write
$$ \frac{B_0}{c} \frac{\dd \Astar}{\dt} = - \left(
-\frac{\va\,B_0}{c} \, \vestar \X \Bstar
-\frac{p_{e0}}{e\,\nave}\,\frac{\grad\pestar}{\nestar} \right)
- \grad\Phi - \frac{c\,B_0}{4\pi}\,\eta \Jstar, $$
and hence, multiplying by $c/(\va\,B_0)$, we obtain 
$$ \frac{1}{\va} \frac{\dd \A}{\dt} = \left(\vestar \X \Bstar
+ \frac{c\,p_{e0}}{\va\,B_0\,e\,\nave} \, \frac{\grad\pestar}{\,\nestar} \right)
- \frac{c}{\va\,B_0} \,\grad\Phi
- \frac{c^2}{4\pi\,\va} \,\eta \Jstar. $$
We thus obtain the normalised Faraday's law
\begin{equation}
\frac{1}{\va} \frac{\dd \Astar}{\dt} = 
- \Estar - \frac{c\,T_{e0}}{\va\,B_0\,e} \, \grad\Phistar,
\label{eqn:normalised_faradays_law}
\end{equation}
where the normalised electric potential $\Phi^*$ is defined as $e\Phi/T_{e0}$ and
the normalised electric field $\Estar$ is defined by a normalised Ohm's law
\begin{equation}
\Estar \equiv \frac{c}{\va\,B_0} \E = \left(- \vestar \X \Bstar
- \frac{c\,T_{e0}}{\va\,B_0\,e} \, \frac{\grad\pestar}{\,\nestar} \right)
+ \eta^* \Jstar,
\label{eqn:normalised_ohms_law}
\end{equation}
the term in parentheses being the ideal part and the remainder the resistive part.

\subsubsection{Evolution of normalised Faraday's law}
\label{sec:b_evolution}

The mean electrostatic potential $\langle \Phi^* \rangle$ is obtained from mean radial
momentum balance. Taking a flux surface average of the covariant $\rho$ component of the 
normalised momentum equation [Eq. (\ref{eqn:normalised_momentum_eqn})], neglecting contributions 
to the pressure gradient term that are nonlinear in flux surface variations of $\nestar$ and 
$\tistar$, and using the fact that the flux surface average of the radial component of $\vi$ must
vanish on turbulent timescales to ensure ambipolarity, we obtain
$$ 0 = - \left \langle \left\{ \left[ \Wstar + \frac{\omega_{ci}}{\va} \Bstar \right]
\X \vistar \right\}_\rho \right \rangle
- \frac{\beta_{i0}}{\langle \nestar \rangle} \frac{d\langle \pistar \rangle}{d\rho} 
- \beta_{e0} \frac{d\langle \Phistar \rangle}{d\rho} 
- \frac{1}{2} \frac{d\langle \vistar^2\rangle}{d\rho}. $$
Rearranging and integrating with respect to $\rho$, we obtain the mean electrostatic potential:
$$ \langle \Phistar \rangle = - {\beta_{i0}\over\beta_{e0}}\int_0^\rho \frac{1}{\langle \nestar 
\rangle} \frac{d \langle \pistar \rangle}{d\rho} d\rho 
- \frac{1}{2\beta_{e0}} \left\{ \langle \vistar^2 \rangle(\rho) - \langle
\vistar^2 \rangle(0) \right\}
 - {1\over\beta_{e0}}\int_0^\rho \left \langle \left\{ \left[ \Wstar + \frac{\omega_{ci}}{\va}
\Bstar \right] \X \vistar \right\}_\rho \right\rangle \, d\rho . $$
In the standard version of \centori\/ we define the total electrostatic potential $\Phistar$ 
using the adiabaticity relation
\begin{equation}
\Phistar = \langle\Phistar\rangle +
\langle\testar\rangle \ln \left( \frac{\nestar}{\langle\nestar\rangle} \right),
\label{eqn:phistar2}
\end{equation}
which follows from electron force balance along the magnetic field in the limit of 
vanishing electron mass \cite{Thyagaraja2006}. We can write  
\begin{equation}
\frac{\dd \Astar}{\dt} = \frac{\dd \tilde{\Astar}}{\dt},
\end{equation}
where $\tilde{\Astar}$ is the fluctuating part of $\Astar$. It follows from Eqs. 
(\ref{eqn:normalised_faradays_law}) and (\ref{eqn:normalised_ohms_law}) that
$$ \frac{1}{\va} \frac{\dd \tilde{\Astar}}{\dt} = 
\vestar \X \Bstar + \frac{c\,T_{e0}}{\va\,B_0\,e} \left( \frac{\grad\pestar}{\nestar} -
\grad\Phistar \right) - \Estar_{\rm res}. $$
The scalar product of this equation with $\Bstar$ yields
\begin{equation}
\frac{1}{\va} \Bstar\cdot\frac{\dd \tilde{\Astar}}{\dt} + \Bstar\cdot\Estar_{\rm res} = 
\frac{c\,T_{e0}}{\va\,B_0\,e} \Bstar\cdot\left(
\frac{\grad\pestar}{\nestar} - \grad\Phistar \right)
\end{equation}
Approximating the time-dependent terms on the left hand side by replacing $\Bstar$ 
with $\Beqstar$, we obtain
$$ \frac{1}{\va} \frac{\dd}{\dt} \left( \Beqstar\cdot\tilde{\Astar} \right)
= \frac{c\,T_{e0}}{\va\,B_0\,e} \left\{ \Beqstar\cdot\left(
\frac{\grad\pestar}{\nestar} - \grad\Phistar \right) + \tilde{\Bstar}\cdot\left(
\frac{\grad\pestar}{\nestar} - \grad\Phistar \right) \right\}
- \Beqstar\cdot\Estar_{\rm res}, $$
where $\tilde{\Bstar}$ is the normalised fluctuating part of the magnetic field.
Neglecting magnetosonic waves (i.e.\ the poloidal component of $\tilde{\Astar}$), the 
covariant representation of $\tilde{\Astar}$ reduces to
$$ \tilde{\Astar} = \tilde{A^*_\zeta} \grad\zeta. $$
In this limit
$$ \tilde{\Bstar} \equiv \curl\tilde{\Astar} = \nabla\times\left(\tilde{A^*_\zeta} 
\grad\zeta\right) = \grad\tilde{A^*_\zeta} \X \grad\zeta, $$
and hence
$$ \frac{1}{\va} \frac{\dd}{\dt} \left( \Beqstar\cdot\tilde{\Astar} \right) = 
\frac{c\,T_{e0}}{\va\,B_0\,e} \left\{
\Beqstar\cdot\left( \frac{\grad\pestar}{\nestar} - \grad\Phistar \right) +
\grad\zeta\cdot\left(
\left( \frac{\grad\pestar}{\nestar} - \grad\Phistar \right) \X
\grad\tilde{A^*_\zeta} \right) \right\}
- \Beqstar\cdot\Estar_{\rm res}. $$
Using the expression for $\Phistar$ [Eq. (\ref{eqn:phistar2})] and the fact that
$\Beqstar\cdot\grad\langle f \rangle = 0$ for any $f$ since $\grad\langle f
\rangle$ is purely radial and $\Beqstar$ has no radial component, we obtain
$$ \Beqstar\cdot\left( \frac{\grad\pestar}{\nestar} - \grad\Phistar \right)
 = \frac{\Beqstar}{\nestar}\cdot\grad(\nestar \tilde{\testar}), $$
and hence, using the fact that $\Beqstar\cdot\tilde{\Astar} = F\tilde{A^*_\zeta}/(B_0R^2)$,
\begin{equation}
\frac{1}{\va} \frac{F}{B_0 R^2} \frac{\dd \tilde{A^*_\zeta}}{\dt} =
\frac{c\,T_{e0}}{\va\,B_0\,e} \left\{
\frac{\Beqstar}{\nestar}\cdot\grad(\nestar \tilde{\testar}) + \grad\zeta\cdot\left[
\left(\frac{\grad\pestar}{\nestar} - \grad\Phistar \right) \X
\grad\tilde{A^*_\zeta} \right] \right\}-\Beqstar\cdot\Estar_{\rm res}.
\end{equation}
We represent $-\Beqstar\cdot\Estar_{\rm res}$ as a diffusion term in this equation by writing
$$ - \Beqstar\cdot\Estar_{\rm res} \simeq \frac{F}{B_0 R^2} R_0\,\jac\,D_\eta
\left( \frac{\dd^2\tilde{A^*_\zeta}}{\drho^2} +
\frac{\langle B_p \rangle^2}{B_p^2} \frac{\dd^2\tilde{A^*_\zeta}}{\dtheta^2}
\right), $$
where the normalised resistive diffusivity is given by
$$ D_\eta(\rho,\theta,\zeta) = \eta^* + \frac{\chi_\eta}{\va}
\left( 1 + q \langle R \rangle^2
\sqrt{\frac{m_i}{m_e}} \left[ f_{JJ}\,\Jfstar^2 + \Wfstar^2
\right] \right), $$
$\chi_\eta$ being a user-defined diffusivity. Turbulent diffusivity terms are present to 
damp out fluctuations occurring at the smallest length scales; these tend to be in the poloidal
direction, close to the magnetic axis. Introducing a parameter $M_A \equiv B_0 R^2/F$ we can 
write
\begin{eqnarray}
\frac{1}{\va} \frac{\dd \tilde{A^*_\zeta}}{\dt} & = &
\frac{c\,T_{e0}}{\va\,B_0\,e} \, M_A \, \left\{
\frac{\Beqstar}{\nestar}.\grad(\nestar \tilde{\testar}) + \grad\zeta. \left[
\left(\frac{\grad\pestar}{\nestar} - \grad\Phistar \right) \X
\grad\tilde{A^*_\zeta} \right] \right\} \nonumber \\
&& + R_0\,\jac\,D_\eta
\left( \frac{\dd^2\tilde{A^*_\zeta}}{\drho^2} +
\frac{\langle B_p \rangle^2}{B_p^2} \frac{\dd^2\tilde{A^*_\zeta}}{\dtheta^2}
\right). \nonumber
\end{eqnarray}
Finally, as in the case of the momentum equation [cf. Eq. (\ref{eqn:numdiff})], we add 
numerical diffusion terms to the right hand side of the normalised Faraday's law.

The above equation is approximated by a finite difference equation which may be 
written in the one-dimensional tridiagonal matrix form
\begin{equation}
\mathcal{A}_{j}\,A^{*\,N+1}_{\zeta\,j-1} +
\mathcal{B}_{j}\,A^{*\,N+1}_{\zeta\,j} + 
\mathcal{C}_{j}\,A^{*\,N+1}_{\zeta\,j+1} = \mathcal{R}_{j},
\end{equation}
where subscripts and superscripts have the same meaning as those in the finite
difference approximation to the momentum equation, the coefficients $\mathcal{A}_j$, $\mathcal{B}_j$
and $\mathcal{C}_j$ do not depend explicitly on $A^*_{\zeta}$, while $\mathcal{R}_j$ depends on the 
latest estimate of this quantity as well as its value at the old timestep and also the latest estimate
and old value of $A^*_{\theta}$. As in the case of the
velocity components, $A^*_\zeta$ at the new time is determined by solving the above tridiagonal 
matrix equation using a predictor-corrector scheme. To ensure that only the fluctuating part is 
actually evolved, the flux surface average of $A^*_\zeta$ is evaluated, and subtracted from
$A^*_\zeta$ to determine $\tilde{A^*_\zeta}$ at the new time.

\subsubsection{Plasma resistivity}

In terms of the flux surface-averaged density, the Spitzer resistivity is given by 
\cite{Spitzer}
\begin{equation}
\eta_{\mbox{\scriptsize Spitzer}}(\rho) = \frac{m_e}{2 e^2 \langle n_e \rangle\tau_{ce}},
\label{eqn:spitzer}
\end{equation}
where 
\begin{equation}
\tau_{ce}(\rho) = \frac{3 \sqrt{m_e} \langle T_e \rangle ^{3/2}}
{4 \sqrt{2\pi} \langle n_e \rangle \lambda e^4},
\label{eqn:tauce}
\end{equation}
is the electron collision time, $\lambda$ being the Coulomb logarithm. In a toroidal 
plasma this is modified by neoclassical effects, which, for singly-charged ions, we model using the 
expression 
\begin{equation}
K_\eta = \frac{1 + \nu_e^*}{(1-\epsilon^{1/2})^2 + \nu_e^*},
\end{equation}
where 
\begin{equation}
\nu_e^* = \frac{\sqrt{2}q R_0}{\epsilon^{3/2}
v_{th,e} \tau_{ce}},
\label{eqn:nuestar}
\end{equation}
is the dimensionless electron collisionality, $q$ being the safety factor of
the flux surface in question, and $\epsilon = a\rho^{1/2}/R_0$ is the local inverse 
aspect ratio. The resistivity is thus
\begin{equation}
\eta(\rho) = K_\eta\eta_{\mbox{\scriptsize Spitzer}(\rho)}.
\label{eqn:resistivity}
\end{equation}
In the banana regime ($\nu_e^* \ll 1$) the above expression for $K_\eta$ 
yields a resistivity which has the appropriate limiting behaviour as $\epsilon 
\to 0$ and $\epsilon \to 1$ \cite{Wesson}; the $\nu_e^*$ dependence ensures that
$K_\eta \to 1$ in the limit of high collisionality, as required.  

\subsubsection{Evolution of toroidal field parameter $F$}
\label{sec:evolve_f}

The loop voltage $V_F$ is related via the resistive MHD form of Ohm's 
law to the part of the toroidal current associated with $FF^\prime$:
$$ \frac{2 V_F}{c\,\eta} = -FF^{\prime}. $$
Defining $V_F^* = V_F/2\pi$, it is straightforward to show that the above equation has the 
following solution for $F$:
$$ F(\rho) = \pm\sqrt{F_{vac}^2 + \frac{8\pi}{c} \psi_0V_F^* \int_\rho^1 
\frac{d\mu}{\eta(\mu)}}, $$
where $F_{vac}$ is the vacuum value of $F$, i.e.\ $F_{vac} = F$ at and outside the
plasma boundary. The plus/minus sign in this expression takes 
into account the possibility of a reversal in the sign of the toroidal magnetic field. 
 
\subsection{Normalised Energy Equations}
\label{sec:normalised_energy_eqns}

We consider here the electron energy equation; the ion equation is treated in
a similar manner. From Eq. (\ref{eqn:energy}) we have
$$ \frac{3}{2} \frac{\partial T_e}{\partial t} + \frac{3}{2} \ve\cdot\grad T_e + T_e 
\nabla\cdot\ve = -\frac{1}{n_e} \nabla\cdot\q_e + \frac{S_e}{n_e}. $$
We can write the first term on the right hand side as
$$ -\frac{1}{n_e} \nabla\cdot\q_e = -\nuparae(T_e - \langle T_e \rangle)
+ \nabla\cdot(X_e \grad T_e), $$
where $X_e$ is the electron thermal conductivity and $\nuparae$ is the parallel electron 
thermal relaxation rate. The latter may be represented by the expression
\begin{equation}
\nuparae = f_{\nuparae} \left( \frac{v_{th,e}}{q \langle R \rangle} \right)
+ \frac{1}{\epsilon^{1/2}\, \tau_{ce}},
\label{eqn:nuparae}
\end{equation}
with $f_{\nuparae}$ a user-defined multiplier and $v_{th,e}(\rho) = (2\langle
T_e \rangle/m_e)^{1/2}$ the electron thermal velocity. This term has the
effect of equilibrating the fluctuating component of $T_e$ rapidly along the
field lines at a rate given by $\nuparae$. The portion involving $X_e$ is
treated as a diffusion term:
$$ \nabla\cdot(X_e \grad T_e) \simeq R_0\,\jac \left\{\left(\chi_e+\chi_{RR}\right) 
\frac{\dd^2 T_e}{\drho^2} 
+ \chi_e \frac{\langle B_p \rangle^2}{B_p^2} \frac{\dd^2 T_e}{\dtheta^2} \right\}. $$
The Rechester-Rosenbluth diffusivity $\chi_{RR}$ can be written as \cite{Rechester}:
\begin{equation}
\chi_{RR} = f_{RR}\,\nuparae \, q^2 \langle R \rangle ^2
\frac{\tilde{B}_{\mbox{normal}}^2}{B^2},
\label{eqn:chirr}
\end{equation}
where $0 \leq f_{RR} \leq 1$ is a user-defined multiplier. Thus the electron
energy equation becomes
\begin{equation}
\frac{3}{2} \frac{\dd T_e}{dt} = -\nuparae T_e + \nuparae \langle T_e
\rangle - T_e \nabla\cdot\ve -\frac{3}{2} \ve\cdot\grad T_e + \frac{S_e}{n_e}
+ R_0\,\jac \left\{
\left( ( \chi_e + \chi_{RR} ) \frac{\dd^2 T_e}{\drho^2} \right)
+ \frac{\langle B_p \rangle^2}{B_p^2} \left( \chi_e \frac{\dd^2
T_e}{\dtheta^2} \right) \right\}.
\end{equation}
The external source term for this equation is
$$ S_e = P_{aux,e} - P_{ei} + \eta J^2, $$
where $P_{aux,e}$ is the external heating power per unit volume provided to
the electrons (see Section~\ref{sec:aux_power}). The second term in the
expression for $S_e$ is the electron-ion equilibration power
[Eq. \ref{eqn:pei})], through which energy is transferred between the
electrons and ions due to the temperature difference between them, and the
final term is the Ohmic heating power. The external source term for the ion
energy equation is
\begin{equation}
S_i = P_{aux,i} + P_{ei},
\end{equation}
where $P_{aux,i}$ is the external heating power per unit volume provided to
the ions, and $P_{ei} = -P_{ie}$ is the electron-ion equilibration power.

We define the following quantities (with the dimensions of reciprocal length):
$$ \nuparae^* \equiv \frac{\nuparae}{\va}, \;\;\;\; D_e \equiv \frac{\chi_e}{\va}, \;\;\;\;
D_{RR} \equiv \frac{\chi_{RR}}{\va}, \;\;\;\;
S_e^* \equiv \frac{S_e}{\va\,T_{e0}\,\nave}. $$
The electron energy equation can then be written in the following normalised form:
\begin{eqnarray}
\frac{1}{\va} \,\frac{\dd \testar}{dt} & = & 
- \frac{2}{3} \nuparae^* \, \testar
+ \frac{2}{3} \nuparae^* \langle \,\testar \rangle
- \frac{2}{3} \,\testar \nabla\cdot\vestar
- \vestar\cdot\grad \testar
+ \frac{2}{3} \frac{S_e^*}{\nestar} \nonumber \\
&& + \frac{2}{3} \,R_0\,\jac \left\{
\left( (D_e + D_{RR}) \frac{\dd^2 \testar}{\drho^2} \right) 
+ \frac{\langle B_p \rangle^2}{B_p^2} \left( D_e \frac{\dd^2 \testar}{\dtheta^2}
\right) \right\}. \nonumber
\end{eqnarray}
In a similar fashion we obtain the normalised ion energy equation:
\begin{eqnarray}
\frac{1}{\va} \,\frac{\dd \tistar}{dt} & = & 
- \frac{2}{3} \nuparai^* \, \tistar
+ \frac{2}{3} \nuparai^* \langle \,\tistar \rangle
- \frac{2}{3} \,\tistar (\nabla\cdot\vistar + \nabla\cdot\Gamma_W^*)
- \vistar\cdot\grad \tistar
+ \frac{2}{3} \frac{S_i^*}{\nestar} \nonumber \\
&& + \frac{2}{3} \,R_0\,\jac\,D_i \left\{
\frac{\dd^2 \tistar}{\drho^2} + \frac{\langle B_p \rangle^2}{B_p^2}
\frac{\dd^2 \tistar}{\dtheta^2} \right\}. \nonumber
\end{eqnarray}
The Ware pinch term $\nabla\cdot\Gamma_W^*$, which is only present in the ion 
equation, is the divergence of the flux \cite{Ware}
\begin{equation}
\Gamma_W^* = -\frac{2.44\epsilon^{1/2}}{\va} \frac{\nestar \,
c}{|B_{\mbox{eq, pol}}|} \, \frac{V_F}{2\pi R} \, \frac{\grad\psi}{|\grad\psi|},
\label{eqn:warepinch}
\end{equation}
and the parallel ion thermal relaxation rate $\nuparai$ is given by
\begin{equation}
\nuparai = f_{\nuparai} \left( \frac{v_{th,i}}{q \langle R \rangle} \right)
+ \frac{1}{\epsilon^{1/2}\tau_{ci}},
\label{eqn:nuparai}
\end{equation}
with $f_{\nuparai}$ a user-defined multiplier. The normalised rate, which again has the 
dimensions of a reciprocal length, is given by $\nuparai^* \equiv \nuparai/\va$.

The normalised electron energy equation is approximated by a
finite difference equation, with the diffusion terms treated exactly by
analogy with those in the momentum equation. This can be written in tridiagonal matrix 
form, and solved at each $(\theta,\zeta)$ point to advance the
normalised electron temperature $\testar$ at the new time. The normalised ion
temperature $\tistar$ is similarly updated.

\subsubsection{Transport of energy}
\label{sec:transport_energy}

The electron collision time is given by Eq. (\ref{eqn:tauce}) and the ion
collision time by the expression
\begin{equation}
\tau_{ci}(\rho) = {3\sqrt{m_i}\langle T_i\rangle^{3/2}\over4\sqrt{\pi}
\langle n_e\rangle\lambda Z_i^4e^4},
\end{equation}
where $Z_i$ is the ion charge state. The power density transferred from
electrons to ions (or vice versa) due to the temperature difference between
them is given by
\begin{equation}
P_{ei}(\rho) = \frac{3 m_e}{m_i} \frac{\langle n_e \rangle}{\tau_{ce}}
\left( \langle T_e \rangle - \langle T_i \rangle \right)
 = \frac{3 m_e}{m_i} \frac{p_{e0}}{\tau_{ce}} \, \langle n_e^* \rangle 
\, \left( \langle T_e^* \rangle - \langle T_i^* \rangle \right).
\label{eqn:pei}
\end{equation}
We define dimensionless collisionalities for the two species by the expressions 
\begin{equation}
\nu_e^*(\rho) = \frac{\sqrt{2} \langle q \rangle R_0}{\epsilon^{3/2}
v_{th,e} \tau_{ce}}, \;\;\;\;\; \nu_i^*(\rho) = \frac{\sqrt{2} \langle q 
\rangle R_0}{\epsilon^{3/2}
v_{th,i} \tau_{ci}},
\end{equation}
where $v_{th,i} = (2\langle T_i\rangle/m_i)^{1/2}$ is the ion thermal
speed. We define flux surface-averaged electron and ion cyclotron frequencies
and thermal Larmor radii by
$$ \omega_{ce}(\rho) = \frac{e \, \langle B \rangle}{m_e c}, \;\;\;\;\;\;
\omega_{ci}(\rho) = \frac{Z_i e \, \langle B \rangle}{m_i c}, \;\;\;\;\;\; 
\rho_e(\rho) = \frac{v_{th,e}}{\omega_{ce}}, \;\;\;\;\;\; 
\rho_i(\rho) = \frac{v_{th,i}}{\omega_{ci}}. $$
We also define poloidal Larmor radii by the expressions
$$ \rho_{pe}(\rho) = \rho_e \, \frac{\langle B \rangle}{\langle B_p \rangle}, \;\;\;\;\;\;
\rho_{pi}(\rho) = \rho_i \, \frac{\langle B \rangle}{\langle B_p \rangle}. $$
The electron and ion neoclassical thermal diffusivities are taken to be
\begin{equation}
\chi_{e,NC}(\rho) = \frac{K_{NC,e} \, \epsilon^{1/2} \rho_{pe}^2}{\tau_{ce}}, \;\;\;\;\;\;
\chi_{i,NC}(\rho) = \frac{K_{NC,i} \, \epsilon^{1/2} \rho_{pi}^2}{\tau_{ci}}, \nonumber
\end{equation}
where $K_{NCi}$ is given by an expression that was proposed by Chang and Hinton
\cite{Chang} as a finite aspect ratio generalisation of a result originally obtained 
by Hinton and Hazeltine \cite{Hinton}
$$ K_{NC,i}(\rho) = \frac{0.66 + 1.88 \epsilon^{1/2} - 1.54 \epsilon}
{1 + \sqrt{\nu_{i}^*} + 0.31 \nu_{i}^*} +
\frac{0.66}{0.31} \frac{\left( (0.74)^2 \epsilon^3 \nu_{i}^* \right)}
{1 + 0.74 \nu_{i}^* \epsilon^{3/2}}, $$
An identical expression is used for $K_{NC,e}$, with $\nu_e^*$ replacing $\nu_i^*$. 
Heat transport in tokamak plasmas is typically found to be due mainly to turbulence 
rather than neoclassical effects, particularly in the case of electrons. In MAST
ion heat transport can be close to neoclassical in the plasma core \cite{Akers}, where the 
approximations used to obtain the above expression for $K_{NC,i}$ are well-satisfied. Closer 
to the plasma edge in MAST, the ion heat transport is generally dominated by turbulence.  

The thermal diffusivities used in the energy equations have the dimensions of length:
\begin{equation}
D_e(\rho,\theta,\zeta) = \frac{1}{\va} \, \left\{ \chi_{e,NC} + \chi_e
\left( 1 + q \langle R \rangle^2
\sqrt{\frac{m_i}{m_e}} \left[ f_{JJ}\,\Jfstar^2 + \Wfstar^2
\right] \right) \right\},
\label{eqn:normdiffse}
\end{equation}
\begin{equation}
D_i(\rho,\theta,\zeta) = \frac{1}{\va} \, \left\{ \chi_{i,NC} + \chi_i
\left( 1 + q \langle R \rangle^2
\sqrt{\frac{m_i}{m_e}} \left[ f_{JJ}\,\Jfstar^2 + \Wfstar^2
\right] \right) \right\},
\label{eqn:normdiffs}
\end{equation}
where $\chi_e$ and $\chi_i$ are background diffusivities specified
by the user. Turbulent diffusivity terms are present in Eqs.~(\ref{eqn:normdiffse}) and 
(\ref{eqn:normdiffs}) to damp out fluctuations occurring at suitably small length scales. 
These model phenomenologically the effect of all fluctuations on subgrid scales, in a manner 
similar to that used in large-eddy simulations in meteorology \cite{Stoll}. In future work 
we intend to derive suitable closure relations by means of kinetic modelling on scales below 
those resolvable using \centori.     

\subsubsection{Auxiliary Heating Power}
\label{sec:aux_power}

There are three options for the auxiliary electron heating power density
profile in \centori:
\begin{equation}
P_{aux,e}(\rho) = \left\{
\begin{array}{cc}
(1-\rho) \, P_{e0} \, e^{-\alpha_{pe} |\rho-\rho_{peak,e}|} & \\
(1-\rho) \, P_{e0} \, e^{-\alpha_{pe} |\rho-\rho_{peak,e}|^2} & \\
P_{e0} \, e^{-\alpha_{pe} |\rho-\rho_{peak,e}|^2} & 
\end{array}
\right. \nonumber
\end{equation}
where $P_{e0}$ gives the height of the profile in erg cm$^{-3}$ s$^{-1}$,
$\alpha_{pe}$ is the profile index, and $\rho_{peak,e}$ is location ($\sim
(r/a)^2$) at which the power profile peaks. These parameters, along with the
choice of profile type, are specified by the user. The ion heating profile is
treated similarly, with an equivalent set of parameters. In principle it is possible to
use profiles obtained from radio-frequency or neutral beam heating codes (applied to
\grass\/ equilibria), and it is essential to do so if precise comparisons with experimental 
results are required.  

\subsection{Normalised mass continuity}
\label{sec:normalised_mass_continuity}

Dividing Eq. (\ref{eqn:continuity_eqn}) by $m_i\,\nave\,\va$, we obtain the normalised mass 
continuity equation
\begin{equation}
\frac{1}{\va} \frac{\dd \nestar}{\dt} =
- \nuparai^* (\nestar - \langle \nestar \rangle)
- \grad\cdot(\nestar\,\vistar)
+ \Snstar - \nabla\cdot\Gamma_W^* + \frac{2}{3} R_0\,\jac \left\{
\left( (D_n + D_{RR}) \frac{\dd^2 \nestar}{\drho^2} \right)
+ \frac{\langle B_p \rangle^2}{B_p^2} \left( D_n \frac{\dd^2
\nestar}{\dtheta^2} \right)\right\},
\label{eqn:normalised_mass_continuity_eqn}
\end{equation}
where the normalised particle source $\Snstar$ (see Section~\ref{sec:particle_source}) is 
given by
\begin{equation}
\Snstar = \frac{S_n}{m_i\,\nave\,\va}
\;\;\; \mbox{(cm$^{-1}$)}.
\label{eqn:snstar}
\end{equation}
The mass continuity equation is approximated by a finite difference equation which, as in the
case of the other primary quantities, can be written in a tridiagonal matrix form suitable 
for advancing in time.   

\subsubsection{Particle source rate}
\label{sec:particle_source}

The rate at which particles (ions) are supplied externally to the plasma per
unit volume is $S_n(\rho)/m_i$. We assume that there are two contributions to
this -- from an auxiliary (neutral beam) power source, if any, and via a
density feedback mechanism (see Section~\ref{sec:density_feedback}). The
latter contribution may be assumed to be highest at the edge, falling to close
to zero at the plasma centre. Thus, the total normalised particle source rate
$\Snstar$ [Eq.~(\ref{eqn:snstar})] is specified in \centori\/ as
\begin{equation}
\Snstar(\rho) \equiv \frac{S_n(\rho)}{m_i\,\nave\,\va} =
\frac{1}{\nave\,\va}\, \left( \frac{P_{aux,i}(\rho)+P_{aux,e}(\rho)}{E_{beam}}
+ S_{\mbox{\scriptsize n edge}} \, C(\rho) \, e^{5(\rho-1)} \right),
\end{equation}
where $S_{\mbox{\scriptsize n edge}}$ is specified in units of
cm$^{-3}$~s$^{-1}$, $P_{aux,i/e}$ is the external heating power provided to
the ions/electrons in ergs cm$^{-3}$ s$^{-1}$, $E_{beam}$ is the neutral beam
particle energy in ergs, and $C(\rho)$ is a cut-off function used to provide
further modulation of the feedback source. Currently we remove the feedback
source completely outside the $\rho^{1/2} = 0.95$ contour, i.e. $C(\rho) = 1$
if $\rho^{1/2} \leq 0.95$ and $C(\rho) = 0$ otherwise.

\subsubsection{Particle diffusion}
\label{sec:particle_diffusion}

We take the normalised particle diffusivity to be related to the normalised
electron thermal diffusivity [Eq.~(\ref{eqn:normdiffs})]:
\begin{equation}
D_n(\rho,\theta,\zeta) = \frac{1}{\va} \, \left\{ \chi_{e,NC} + \chi_{ne}
\left( 1 + q \langle R \rangle^2
\sqrt{\frac{m_i}{m_e}} \left[ f_{JJ}\,\Jfstar^2 + \Wfstar^2
\right] \right) \right\},
\end{equation}
where $\chi_{ne}$ is a user-defined particle diffusivity. As in the case of $\chi_e$ and 
$\chi_i$ in the thermal diffusivity expressions [Eqs.~(\ref{eqn:normdiffse}) and 
(\ref{eqn:normdiffs})], this is used to model transport arising from processes occurring on 
sub-grid scales; typically $\chi_{ne} \sim 10^4\,$cm$^2$s$^{-1}$.

\subsubsection{Density feedback}
\label{sec:density_feedback}

There is an option in \centori\/ to use a feedback mechanism to control the
volume-averaged particle density. This is achieved by modifying the edge
particle source rate $S_{\mbox{\scriptsize n edge}}$ at each timestep as
follows:
\begin{equation}
S_{\mbox{\scriptsize n edge}} = \left\{
\begin{array}{cl}
\left(n_{\mbox{\scriptsize e target}} - N_{\mbox{\scriptsize total}}/V\right)/\tau_{sn} 
& \mbox{if} \;\;\; n_{\mbox{\scriptsize e target}} >
N_{\mbox{\scriptsize total}}/V \\
0 & \mbox{otherwise}
\end{array} \right.,
\end{equation}
where $n_{\mbox{\scriptsize e target}}$ is the requested average density,
$N_{\mbox{\scriptsize total}}$ is the total number of particles in the plasma
(i.e.\ the volume integral of $n_e$), $V$ is the plasma volume, and
$\tau_{sn}$ is the required timescale for the density to reach the target
value. If the density is too high the particle source is turned off.

\section{Initial and boundary conditions}
\label{sec:boundary_conditions}

\subsection{Initial conditions}

At $t=0$ the physical quantities are prescribed as follows. All
fluctuating components are initialised to zero, except for $\tilde{n_e}$,
which is given an arbitrary variation in all three directions.
$$ v_{i,\,\mbox{normal}}(\rho,\theta,\zeta) =
 v_{i,\,\mbox{tangential}}(\rho,\theta,\zeta) = 0, \;\;\;\;\;
 v_{i,\,\mbox{toroidal}}(\rho,\theta,\zeta) = v^*_{i0}\,\va\,e^{-\alpha_{vi}\rho}, $$
$$ n_e(\rho,\theta,\zeta) = n_{e0} \, e^{-\alpha_n \rho}, \;\;\;\;\;
T_e(\rho,\theta,\zeta) = T_{e0} \, e^{-\alpha_{te} \rho}, \;\;\;\;\;
 T_i(\rho,\theta,\zeta) = T_{i0} \, e^{-\alpha_{ti} \rho}, $$
The coefficients and profile indices in the above expressions are specified by the user. 
The initial vector potential $\A$ and magnetic field $\B$ are derived from the
initial equilibrium $\psi(R,Z)$, as described in Section~\ref{sec:coordinates}. In the early
stages of a simulation it may be necessary to determine an equilibrium relatively frequently
(typically once every 10$^3$ time steps) to allow transients to settle.
This early-stage evolution does not simulate accurately the startup phase of a real plasma.   

\subsection{Boundary conditions}

The boundary conditions in the $\theta$ and $\zeta$ directions are, of course,
periodic.  In this section we discuss the boundary conditions to be applied in
the radial direction.

\subsubsection{Axis boundary conditions}

At each discrete toroidal location $\zeta_n$ the true plasma axis
$(\rho=0,\theta,\zeta=\zeta_n)$ is a coordinate singularity, since $\theta$ is
undefined (i.e.\ it can take any value from 0 to $2\pi$). The radial and
poloidal directions are similarly undefined. There is still a clearly-defined
toroidal direction, however. With these considerations in mind, the physical
components of all vector quantities at the plasma axis are dealt with as
follows. If $\mathbf{V}(\rho,\theta,\zeta)$ denotes any vector quantity, and
$\rho = \Delta\rho$ denotes the radial location of the first grid point away
from the axis, then the normal and toroidal vector components are given by
$$ V_{\mbox{normal}}(0,\theta,\zeta_n) = \mbox{mean value of }V_{\mbox{normal}}
(\Delta\rho,\theta,\zeta_n), $$
$$ V_{\mbox{toroidal}}(0,\theta,\zeta_n) = \mbox{mean value of }V_{\mbox{toroidal}}
(\Delta\rho,\theta,\zeta_n), $$
while the tangential component is set equal to zero. As previously noted, the
value of $\rho$ closest to the axis has a small positive value. The scalar
quantities $n_e$, $T_e$, $T_i$, $p_e$ and $p_i$ are treated in the same way as
$V_{\mbox{normal}}$ and $V_{\mbox{toroidal}}$, while the flux surface-averaged
profiles of these quantities are assumed to be flat at the magnetic axis. In
the case of the density profile, for example,
$$ \langle n_e \rangle (0) = \langle n_e \rangle (\Delta\rho). $$
Similar boundary conditions are applied at the axis to $\langle T_e \rangle$,
$\langle T_i \rangle$, $\langle p_e \rangle$, $\langle p_e \rangle$ and
$\langle p_i \rangle$.

\subsubsection{Edge boundary conditions}

The edge of the plasma is less problematic in terms of the coordinate system
than the axis. All four of the following boundary conditions are used for
different quantities $f$ in the code:
\begin{itemize}

\item \textbf{Zero}: $f(\rho=1,\theta,\zeta) = 0$.

\item \textbf{Flat gradient}: $\dd f/\drho = 0$, i.e.\ $f(\rho=1,\theta,\zeta)
= f(\rho=1-\Delta\rho,\theta,\zeta)$.

\item \textbf{Continuous gradient}: $\dd f/\drho$ is constant, i.e. 
$\partial^2f/\partial\rho^2 = 0$:
\begin{equation}
f(\rho=1,\theta,\zeta) = 2 \, f(\rho=1-\Delta\rho,\theta,\zeta) -
f(\rho=1-2\Delta\rho,\theta,\zeta) \nonumber
\end{equation}

\item \textbf{Fixed}: $f(\rho=1,\theta,\zeta)$ is held fixed at some
predetermined value.

\end{itemize}
These boundary conditions are applied as shown in Table~\ref{tab:edgebcs}.

\begin{table}[ht]
\begin{center}
\begin{tabular}{|c|l|} \hline
\textbf{quantity} & \textbf{edge boundary condition} \\ \hline
 $\tilde{A_\zeta}$, $\W$, 
	 $\langle v_{i,\mbox{normal}}\rangle$ & zero \\
 $\B$ & (contravariant) continuous gradient \\
 $\J$ & (contravariant) flat gradient \\
 $\vi$ & (physical) flat gradient (but toroidal component zero) \\
 $\ve$ & (physical) flat gradient \\
 $n_e$, $T_e$, $T_i$ & fixed \\
 $\Phi$, $\langle n_e \rangle$,  $\langle T_e \rangle$, 
	$\langle T_i \rangle$  & continuous gradient \\
 $\langle v_{i,\mbox{tangential}}\rangle$, $\langle v_{i,\mbox{toroidal}}\rangle$ 
	& continuous gradient \\ \hline
\end{tabular}
\caption{\label{tab:edgebcs} Plasma edge boundary conditions applied to evolving quantities 
in \centori.}
\end{center}
\end{table}

\section{Evolution of mean and fluctuating components}
\label{sec:evolution}

\subsection{Scalar quantities}
\label{sec:meanscalars}

In Section~\ref{sec:normalised_physics_eqns} we discussed the equations
governing the evolution of physics quantities in \centori. Each of these
quantities can be split into mean (or equilibrium) and fluctuating parts. The
``mean'' of a scalar quantity $f$ in this context simply refers to its flux
surface average, as defined by Eq.~(\ref{eqn:fluxsurfaceaverage}), and the
fluctuating component $\tilde{f}$ is the remainder:
$$ f_{\mbox{\scriptsize total}} = \langle f \rangle + \tilde{f}. $$
In the case of normalised electron density, for example, we have
$$ \nestar(\rho,\theta,\zeta) = \langle \nestar \rangle (\rho) +
\tilde{\nestar}(\rho,\theta,\zeta). $$
The normalised quantities $\testar$, $\tistar$, $\pestar$ and $\pistar$ are
split in a similar fashion. In each case the flux surface average is evaluated
at each timestep after the total quantity has been updated, and the
fluctuating component is obtained simply by subtracting the average from the
total.

\subsection{Vector quantities}

The fluctuating components of vector quantities are obtained in a similar
fashion by subtraction of means from totals, but the means themselves are
calculated differently.  The physical components of the mean ion velocity
$\vieq$ are given by the flux surface averages of the corresponding components
of the total ion velocity $\vi$. The mean electron velocity $\veeq$, on the
other hand, is obtained from $\vieq$ and $\Jeq$ using the flux
surface-averaged form of Eq.~(\ref{eqn:vestar}).

The electromagnetic equilibrium vector quantities only need to be re-evaluated
when the plasma equilibrium is updated (see
Section~\ref{sec:equilibrium_solver}), i.e.\  when $\psi(R,Z)$ is
recalculated. Then, the mean vector potential $\Aeq$ is determined using
Eqs.~(\ref{eqn:arho}),~(\ref{eqn:azeta}) and~(\ref{eqn:atheta}). The mean
magnetic field $\Beq$ is obtained directly from the curl of $\Aeq$, and the
mean current density $\Jeq$ is obtained from $\Beq$ via Amp\`ere's law
[Eq.~(\ref{eqn:amperes_law})]. However, Eq.~(\ref{eqn:atheta}) shows that the
covariant $\theta$ component of $\Aeq$ depends on $F(\psi)$, which determines
the toroidal magnetic field [cf. Eq.~(\ref{eqn:beq})]. The evolution of $F$ is
described in Section~\ref{sec:evolve_f}.

\section{Global energy-related quantities}
\label{sec:globals}

The Ohmic heating power density is
$$ P_{ohm} = \eta \, \J^2 = \frac{\va\,B_0^2}{4\pi} \, \eta^* \, \Jstar^2. $$
The kinetic energy densities in the electrons and ions are given by 
$$ E_{k,e} = \frac{1}{2} \int m_e \,\nave \, \va^2 \, \nestar \, (\vestar\cdot\vestar) \; dV, 
\;\;\;\;\;\; E_{k,i} = \frac{1}{2} \int m_i \,\nave \, \va^2 \, \nestar \, (\vistar\cdot\vistar) 
\; dV. $$
The total thermal energy and magnetic field energy are
$$ E_{th} = \frac{3}{2} \int p \; dV, \;\;\;\;\;\; E_B = \frac{1}{8\pi} \int B^2 \; dV, $$
where $p$ is the total pressure. We define the total plasma beta as
$$ \beta = 8\pi\frac{\int p \; dV}{\int B^2 \; dV}
      = \frac{2}{3} \frac{E_{th}}{E_B}. $$
Similarly, the poloidal beta is defined to be
$$ \beta_p = \frac{8\pi\int p \; dV}{\int B_p^2 \; dV}
= \frac{16\pi E_{th}}{3\int B_p^2 \; dV}. $$

\section{Equilibrium force balance and the Grad-Shafranov equation}
\label{sec:equilibrium_solver}

The Grad-Shafranov equation, which can  be derived from the steady-state form 
of the two-fluid equations \cite{Thyagaraja2006}, describes the equilibrium state of a
current-carrying magnetised plasma in which the Lorentz force is balanced by a
pressure gradient force. As described below, a pseudo-transient approach is used in  
\centori\/ to solve this equation. Similar techniques have been employed in computational 
fluid dynamics \cite{Fletcher}, but have not, as far we are aware, been applied previously 
to the problem of determining toroidal plasma equilibria. The Grad-Shafranov equation can be 
generalised to include transonic flows and momentum sources \cite{McClements2011}. 
Currently, however, only the simplest form of the equation, which is applicable 
when toroidal flows are subsonic and poloidal flows are less than the sound speed 
multiplied by the ratio of the poloidal magnetic field to the total field \cite{Hameiri}, 
is used in \centori\/; it can be written in the form
\begin{equation}
R{\partial\over\partial R}\left({1\over R}{\partial\psi\over\partial R}\right)
+{\partial^2\psi\over\partial Z^2} \equiv \delpsi = -4\pi R^2 p^\prime - F F^\prime,
\label{eqn:grad_shafranov}
\end{equation}
where primes denote derivatives with respect to $\psi$. The equation can also be written
in the form
\begin{equation}
\frac{4\pi}{c} J_\zeta = -4\pi R^2 p^\prime - F F^\prime,
\label{eqn:grad_shafranov2}
\end{equation}
where $J_\zeta = (c/4\pi) \, \delpsi$ is the covariant $\zeta$ component of
the equilibrium current density, $\Jeq$. Although flow 
modifications to equilibrium flux surfaces are neglected in the current version of \centori\/,
the effects of low Mach number flows and flow shear on turbulence and MHD instabilities 
are taken into account in the two-fluid equations described in 
Section~\ref{sec:twofluidequations}. Thus, \centori\/ can be used to model, amongst other things,
the stabilising effects of sheared flows on ion temperature gradient modes \cite{Migliuolo} and the 
destabilising effects of such flows on Kelvin-Helmholtz instabilities \cite{Chapman}. It is anticipated 
that flow effects on plasma equilibria will be taken into account in future versions of the code; 
users of the present version should note that it is strictly applicable only to subsonic equilibrium flows.

\subsection{The \grass\/ free boundary equilibrium solver}
\label{sec:grass}

\begin{figure}[ht]
\begin{center}
\epsfig{file=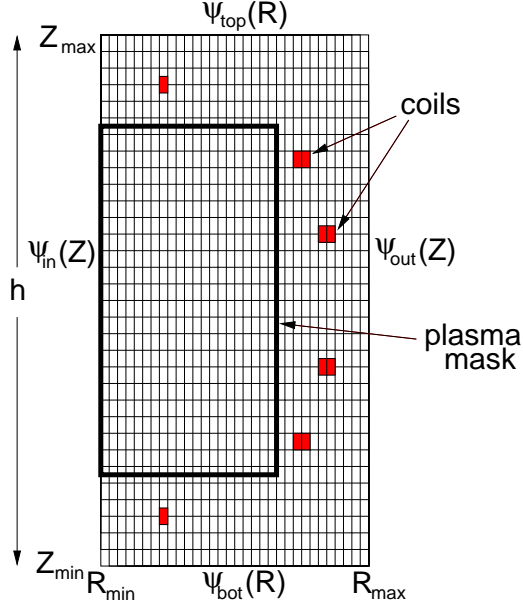,height=8cm,angle=0}
\parbox{14cm}{
\caption{\label{fig:grassgrid} Schematic diagram of the computational domain used in the \grass\/
equilibrium solver, showing the main solution grid and the plasma mask.}  }
\end{center}
\end{figure}

The \centori\/ source code includes a free boundary Grad Shafranov equilibrium
solver named \grass\/ \cite{ThyagarajaKnight} (GRAd Shafranov Solver), which is used 
to compute solutions of Eq.~(\ref{eqn:grad_shafranov}), taking into account the presence
of currents in poloidal field coils. Figure~\ref{fig:grassgrid} shows the
layout of the computational domain used in this subroutine. The toroidal field coils
are assumed to lie entirely outside the computational domain; as described in 
Section~\ref{sec:evolve_f}, the toroidal field parameter $F$ is determined by the loop
voltage and the resistivity. 
The solver uses two rectangular grids:
\begin{enumerate}
\item The main solution grid, within the domain $(R_{\rm min},Z_{\rm min})$ to
  $(R_{\rm max},Z_{\rm max})$. The plasma and the coils are assumed to lie
  wholly within this grid. The poloidal flux values on the grid boundaries
  $\psi_{\rm in}$, $\psi_{\rm out}$, $\psi_{\rm top}$ and $\psi_{\rm bot}$ are
  calculated analytically from the given coil currents and an approximation to
  the current distribution in the plasma region.
\item The plasma mask, comprising the rectangular region $(R_{\rm p\,min},Z_{\rm p\,min})$
  to $(R_{\rm p\,max},Z_{\rm p\,max})$. The mask must not extend outside the
  main solution grid. The (hot) plasma is assumed to lie wholly within the
  plasma mask, but no coils can be present inside it.
\end{enumerate}
The coils' current density $J_c$ (which needn't be the same in each coil) is
assigned to a number of grid cells, to approximate the coil locations and
cross-section areas.

\subsubsection{\grass\/ solution procedure}

It is necessary to solve the following equation over the main solution grid:
$$ \delpsi = \frac{4\pi}{c} R J_{tor}, $$
where $J_{tor}$ is a function of $\psi$ throughout the region containing plasma, and
$J_{tor} = J_c$ at the coil locations. Thus we can rewrite the equation as
\begin{equation}
\delpsi = \frac{4\pi}{c} \left( R J_t\,H + R J_c \right) \label{eqn:gs},
\end{equation}
where
\begin{equation}
H = \left\{ \begin{array}{rl}
1 & \mbox{inside plasma mask} \\
0 & \mbox{elsewhere} 
\end{array} \right. \nonumber
\end{equation}
and $J_t$ is the toroidal component of $\Jeq$ within the plasma:
\begin{equation}
R J_t = -c\,R^2 p^\prime - \frac{c}{4\pi} F F^\prime .
\label{eqn:rjtor}
\end{equation}
We denote by $E_F$ the toroidal electric field that drives the portion of the toroidal
current density proportional to $-F F^\prime$. From the resistive MHD form of
Ohm's law we thus have
$$ \frac{4\pi}{c} R J_t = -4\pi\,R^2 p^\prime + \frac{4\pi R}{c} \frac{E_F}{\eta}. $$
Setting $2\pi R\,E_F \equiv V_F$, the equivalent loop voltage, we obtain
\begin{equation}
RJ_t = -\frac{c\,R^2}{\Delta\psi} \frac{dp}{d\rho} + \frac{V_F^*}{\eta},
\label{eqn:newRJ}
\end{equation}
where $V_F^* = V_F/2\pi$ and $\Delta\psi = \psi_{\mbox{\scriptsize{edge}}} - 
\psi_{\mbox{\scriptsize{min}}}$ (see below). The dependencies of $dp/d\rho$ 
and $\eta$ on $\psi$ are prescribed.

To determine $V_F^*$ we divide Eq.~(\ref{eqn:newRJ}) by $R$ and integrate over
the poloidal cross-section area, identifying this quantity as the total plasma
current $I_p$:
$$ I_p \equiv \int J_t\,dA = -\int \frac{c\,R}{\Delta\psi}
\frac{dp}{d\rho}\,dA + \int \frac{V_F^*}{R\,\eta}\,dA = -\frac{c}{\Delta\psi} \int R 
\frac{dp}{d\rho}\,dA + V_F^* \int\frac{dA}{R\,\eta}. $$
It follows from this that
$$ V_F^* = \frac{I_p - \left( -\frac{c}{\Delta\psi} \int R
\frac{dp}{d\rho}\,dA \right)}{\int \frac{dA}{R\,\eta}}. $$
It is convenient to introduce a new dependent variable $u = \psi/R^{1/2}$
satisfying the boundary conditions
$$ u_{\rm in}(Z) = \frac{\psi_{\rm in}(Z)}{R_{\rm min}^{1/2}}, \;\;\;
u_{\rm out}(Z) = \frac{\psi_{\rm out}(Z)}{R_{\rm max}^{1/2}}, $$
$$ u_{\rm bot}(R) = \frac{\psi_{\rm bot}(R)}{R^{1/2}}, \;\;\;
u_{\rm top}(R) = \frac{\psi_{\rm top}(R)}{R^{\frac{1}{2}}}. $$
It is also convenient to express $\psi$ as the sum of two terms: $\psi_1$,
which vanishes at $Z=Z_{\rm min}$ and $Z = Z_{\rm max}$; and
$$ \psi_2 \equiv \frac{Z-Z_{\rm min}}{h} \psi_{\rm top} + 
\frac{Z_{\rm max}-Z}{h}\psi_{\rm bot}, $$
where $h = Z_{\rm max}-Z_{\rm min}$. The quantity $\psi_1$ is then equal to 
$\psi-\psi_2$. Equivalently,
$$ u_2 = \frac{Z-Z_{\rm min}}{h} u_{\rm top} + \frac{Z_{\rm max}-Z}{h} u_{\rm bot}, $$
and $u_1 \equiv u - u_2$. Clearly $u_1$ vanishes at $Z=Z_{\rm min}$ and $Z=Z_{\rm
  max}$, making it possible to compute this quantity by applying a sine
Fourier transform in $Z$.

Defining the operator $\Delta^*_u$ by the equation
$$ \Delta^*_u u \equiv \frac{1}{R^{1/2}} \delpsi = 
\frac{\dd^2 u}{\dd R^2} + \frac{\dd^2 u}{\dd Z^2} - \frac{3}{4R^2} u, $$
we find that the Grad-Shafranov equation becomes
\begin{equation}
\Delta^*_u u_1 = 
\frac{\dd^2 u_1}{\dd R^2} + \frac{\dd^2 u_1}{\dd Z^2} - \frac{3}{4R^2} u_1 =
\frac{4\pi}{c} \left( \frac{R J_t\,H}{R^{1/2}} + R^{1/2}
J_c \right) - \Delta^*_u u_2.
\label{eqn:tmp06}
\end{equation}
Since $u_2$ is a prescribed function of $Z$, the quantity $\Delta^*_u u_2$
only needs to be evaluated once, at the beginning of the calculation. Moreover
we can set $\Delta^*_u u_2 = -3u_2/(4R^2)$, since it is independent of $R$
and depends only linearly on $Z$.

We approach the problem of solving Eq.~(\ref{eqn:tmp06}) by considering the
parabolic equation
\begin{equation}
\frac{\dd u_1}{\dd\tau} = \epsilon \left( \Delta^*_u u_1 - \alpha \right)
\label{eqn:heateqn}
\end{equation}
where $\alpha$ is the right hand side of Eq.~(\ref{eqn:tmp06}), $\epsilon$ is
a prescribed pseudo-conductivity (taken to be uniform across the poloidal plane)
and $\tau$ is a fictitious, time-like iteration variable (not to be confused with the 
true time, $t$). The problem of solving Eq.~(\ref{eqn:tmp06}) 
is thus equivalent to finding ``steady-state'' solutions of Eq.~(\ref{eqn:heateqn}).    
The sine transform of Eq.~(\ref{eqn:heateqn}) can be
approximated by the finite difference equation
$$ \frac{\hat{u}^{N+1}_{1\,i} - \hat{u}^{N}_{1\,i}}{\Delta\tau} =
\epsilon \left[
\frac{\hat{u}^{N+1}_{1\,i+1} - \hat{u}^{N+1}_{1\,i}}{(\Delta R)^2} -
\frac{\hat{u}^{N+1}_{1\,i} - \hat{u}^{N+1}_{1\,i-1}}{(\Delta R)^2}
- \frac{\pi^2 k^2_Z}{h^2} \hat{u}^{N+1}_{1\,i}
- \frac{3}{4 R^2_i} \hat{u}^{N+1}_{1\,i} - \widehat{\alpha_i} \right], $$
where the $k_Z$-th sine transform coefficients are denoted by
$\widehat{\ldots}$, $i$ labels the $i$-th grid point in the $R$ direction,
with grid spacing $\Delta R$, $N$ labels the pseudo-time variable, and
$\Delta\tau$ is the pseudo-time step. This equation can be rearranged to give
$$ \hat{u}^{N+1}_{1\,i} \left( 1 + \Delta\tau\epsilon
\left\{ \frac{2}{(\Delta R)^2} + \frac{\pi^2 k^2_Z}{h^2} + \frac{3}{4 R^2_i}
\right\} \right)
- \frac{\Delta\tau\epsilon}{(\Delta R)^2} \hat{u}^{N+1}_{1\,i+1}
- \frac{\Delta\tau\epsilon}{(\Delta R)^2} \hat{u}^{N+1}_{1\,i-1} =
\hat{u}^{N}_{1\,i} - \Delta\tau\epsilon \, \widehat{\alpha_i}, $$
which is a tridiagonal matrix equation of the form
$$ \mathcal{A}_i \, \hat{u}^{N+1}_{1\,i-1} + \mathcal{B}_i \, \hat{u}^{N+1}_{1\,i} +
\mathcal{C}_i \, \hat{u}^{N+1}_{1\,i+1} = 
\left( \hat{u}^{N}_{1\,i} - \Delta\tau\epsilon \, \widehat{\alpha_i} \right) $$
where
$$ \mathcal{A}_i = - \frac{\Delta\tau\epsilon}{(\Delta R)^2}, \;\;\;\;\; \mathcal{B}_i =  
1 + \Delta\tau\epsilon
\left\{ \frac{2}{(\Delta R)^2} + \frac{\pi^2 k^2_Z}{h^2} + \frac{3}{4 R^2_i}
\right\}, \;\;\;\;\; \mathcal{C}_i = - \frac{\Delta\tau\epsilon}{(\Delta R)^2}. $$
The tridiagonal matrix equation is straightforward to solve for
$\hat{u}^{N+1}_1$; the inverse sine transform of this yields $u_1$ and thereby
$\psi_1$. The total flux $\psi$ is recovered by adding $\psi_2$, and the
process is repeated until $\psi$ over the grid does not change significantly
between pseudo-time steps.

\subsubsection{Plasma current}

In general the plasma current density $J_t$ and $\Delta\psi$ change
between successive pseudo-time steps, and so the the evolution described by
Eq.~(\ref{eqn:heateqn}) is non-linear. It should be noted that the
dependencies of $dp/d\rho$ and $\eta$ on $\psi$ (or $\rho$) are determined
externally using \centori, rather than \grass. Ideally these functions should
vary with $\psi$ in such a way that the residual plasma current outside of the
chosen edge plasma contour remains negligible.

\subsubsection{Defining the plasma edge}

Once a convergent solution for the equilibrium has been obtained, it is
necessary to locate the edge of the plasma, which is defined to lie wholly
within the plasma mask. By estimating $|\grad\psi|$ at all grid points within
the mask using finite differences, it is straightforward to find all the
stationary points of $\psi$; these are either X-points (saddle points) or the
magnetic axis, which is defined to lie at the global minimum of $\psi$ within
the mask. There should be no other stationary points of $\psi$ inside the
mask, unless some coils have been erroneously located within it.  The edge of
the plasma is then defined after finding a reference $\psi_{\rm max}$ using
the following criteria (the situation is topologically more
complicated in general, but in practice this algorithm suffices):
\begin{itemize}
\item If there are no X-points within the mask, $\psi_{\rm max}$ is chosen to be
the minimum value of $\psi$ along the plasma mask perimeter or the minimum
value of $\psi$ at a user-defined set of $(R,Z)$ ``limiter'' points within the
mask, whichever is lower.
\item If any X-points are present, $\psi_{\rm max}$ is chosen to be
either the $\psi$ of the lowest X-point or the minimum
value of $\psi$ along the inner or outer edges of the mask or the limiter
points, if this is lower than the $\psi$ of the lowest X-point.
\end{itemize}
This ensures that the $\psi_{\rm max}$ contour is the largest closed
contour within the mask. We then define the plasma edge contour $\psi_{\rm edge}$
to be
$$ \psi_{\rm edge} = \psi_{\rm axis} + f(\psi_{\rm max}-\psi_{\rm axis}), $$
where $f = 0.99$ when no X-points are present within the mask and $f = 0.90$
otherwise. This has the effect of moving the effective plasma boundary to a
contour lying slightly inside the last closed flux surface, which is necessary
to ensure that the coordinate system described in Section~\ref{sec:coordinates}
does not become strongly distorted near the plasma edge, and enables us to 
approximate the physics equations with central differences without incurring 
unacceptably large truncation errors.  

Finally, $\psi$ is redefined within the plasma mask so
that the plasma edge corresponds to $\psi=0$. \centori\/ is passed only this
modified $\psi(R,Z)$ within the masked region (thus excluding the coils),
interpolated onto a grid of the same size (i.e.\ with the same number of
elements) using Chebyshev fits in $R$ and $Z$. The algorithm for determining
plasma-based coordinates described in Section~\ref{sec:coordinates} works
extremely well when $\psi(R,Z)$ is specified in this way, and almost
invariably yields a Jacobian $\jac$ that closely approximates a flux function
as a result.

\subsubsection{Control of the magnetic axis location}
\label{sec:position_control}

It is sometimes useful to be able to hold the magnetic axis at a specified
$(R,Z)$ position. For example, up-down asymmetric plasmas are often vertically
unstable, and in such cases it may be difficult to obtain a convergent
solution for the equilibrium using \grass\/ unless it is possible to control
the plasma location during the convergence cycle.

Applying a vertical magnetic field makes it possible to control the
radial position of the plasma, as follows. Suppose that there is a
source of poloidal flux $\psi_{BZ}$ of the form
$$ \psi_{BZ} = \psi_{BZ0} \, \frac{R^2}{R_0^2}, $$
where $\psi_{BZ0}$ is a constant and $R_0$ is a measure of the major radius
(e.g.the value of $R$ at the centre of the computational grid). Then
$$ \grad\psi_{BZ} = \frac{\dd \psi_{BZ}}{\dd R} \er = 2R \frac{\psi_{BZ0}}{R_0^2} \er . $$
Since the poloidal magnetic field is $\grad\zeta \X \grad\psi$ it follows that
the field due to $\psi_{BZ}$ is uniform and vertical:
$$ \B_Z = \frac{2\psi_{BZ0}}{R_0^2} \ez. $$
The Lorentz force $\J_{\rm plas} \X \B_Z$ on the plasma arising from a
positive plasma current $\J_{\rm plas} = J_{\rm plas}\,\ezeta$ will be inwards
if $\psi_{BZ0} > 0$.

Similarly, an externally-provided radial magnetic field affects the plasma's
vertical position. The radial field due to a poloidal flux of the form
$$ \psi_{BR} = \psi_{BR0} \, \frac{Z-Z_0}{Z_{\rm max}-Z_{\rm min}}, $$
is
$$ \B_R = \frac{- \psi_{BR0}}{R(Z_{\rm max}-Z_{\rm min})} \er, $$
and the Lorentz force on the plasma in this case is downwards if $\psi_{BR0} > 0$ and 
$J_{\rm plas} > 0$. 

We adjust the values of $\psi_{BR0}$ and $\psi_{BZ0}$ during each \grass\/
convergence step by comparing the latest calculated position of the magnetic
axis ($R_{\rm axis},Z_{\rm axis}$) with the target location 
($R_{\rm target},Z_{\rm target}$), and applying a correction 
to the fluxes as follows:
$$ \psi_{BR0} \rightarrow \psi_{BR0} +
f\,\psi_0\,\frac{Z_{\rm axis}-Z_{\rm target}}{Z_{\rm max}-Z_{\rm min}}, $$
$$ \psi_{BZ0} \rightarrow \psi_{BZ0} +
f\,\psi_0\,\frac{R_{\rm axis}-R_{\rm target}}{R_{target}}, $$
where $f \ll 1$ (typically $f \sim 0.02$) to ensure that the change in the
fluxes is not substantial. At $t=0$ we assume that $\psi_{BR0} = \psi_{BZ0} =
0$. The applied corrections should modify the radial and vertical fields 
in such a way that the magnetic axis is pushed towards the target location.

It is important to note that the magnetic fields associated with these externally-applied 
poloidal flux components are curl-free and hence current-free, i.e.\ there are no 
additional current sources within the grid implied by their presence.
Experimentally, vertical and radial magnetic field perturbations of this type can be 
introduced by changing the currents in poloidal field coils, although such field perturbations
are in general non-uniform and therefore the uniform field perturbations discussed here are 
somewhat idealised. It is straightforward to incorporate the additional fluxes into \grass\/ by 
simply modifying the boundary conditions at the edge of the computational grid, at
the start of each convergence loop, as follows:
$$ u_{\rm in}(Z) = R_{\rm min}^{-1/2} \left( \psi_{\rm in}(Z) + 
\psi_{BR0} \frac{Z-Z_0}{Z_{\rm max}-Z_{\rm min}} +
\psi_{BZ0} \frac{R_{\rm min}^2}{R_0^2} \right), $$
$$ u_{\rm out}(Z) = R_{\rm max}^{-1/2} \left( \psi_{\rm out}(Z) +
\psi_{BR0} \frac{Z-Z_0}{Z_{\rm max}-Z_{\rm min}} +
\psi_{BZ0} \frac{R_{\rm max}^2}{R_0^2} \right), $$
$$ u_{\rm bot}(R) = R^{-1/2} \left( \psi_{\rm bot}(R) +
\psi_{BR0} \frac{Z_{\rm min}-Z_0}{Z_{\rm max}-Z_{\rm min}} +
\psi_{BZ0} \frac{R^2}{R_0^2} \right), $$
$$ u_{\rm top}(R) = R^{-1/2} \left( \psi_{\rm top}(R) +
\psi_{BR0} \frac{Z_{\rm max}-Z_0}{Z_{\rm max}-Z_{\rm min}} +
\psi_{BZ0} \frac{R^2}{R_0^2} \right). $$

\section{Outline of code structure}
\label{sec:code_outline}

\subsection{Source files}

The \centori\ code is written in standard Fortran 95 throughout, and is contained within
some 21 source files. The bulk of these contain utility modules and routines
to perform specific tasks such as I/O, parallel (MPI) communication, error
handling, numerical evaluation (Chebyshev/Fourier fitting, and so on) and other
customised but standard functionality. To make the code as portable as possible, 
we have avoided the use of external numerical libraries. Those areas of the
code in which such libraries might improve performance almost all occur
in non-parallel segments, i.e. are run only by the ``global''
processor. Since the execution of the code is  overwhelmingly dominated by periods of parallel 
execution, serial optimisation through the use of specialised libraries is
unlikely to confer significant benefits.

The physics within the code described in this paper is confined to two
source files. One of these contains
all the routines for initialising and evolving the physical fields, sources, sinks and 
so on, and also the routines for calculating plasma coordinates
from the $\psi(R,Z)$ grid. The other contains the \grass\ free boundary
equilibrium solver as described in the previous section.

\subsection{Parallelization model}

\begin{figure}[ht]
\begin{center}
\psfig{file=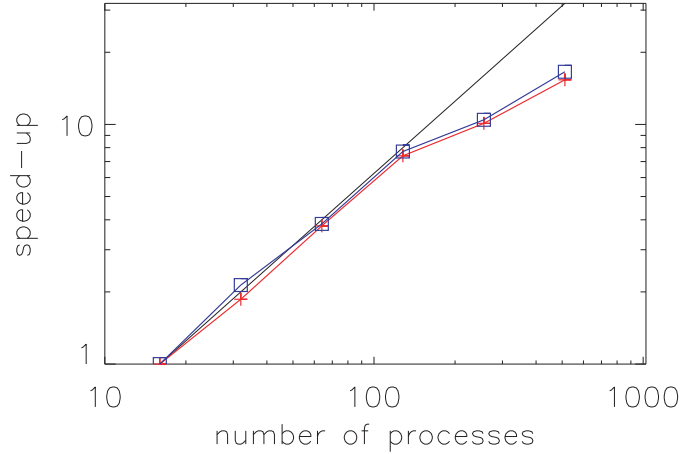,height=6cm,angle=0}
\parbox{14cm}{ \caption{\label{fig:speedup} Execution time speed-up versus process count on 
HECToR (red with $+$ symbols) and HPC-FF (blue with $\Box$ symbols), relative to the time taken 
for a 16-process run, for a computational grid of $129 \times 65 \times 33$. This is the typical grid used for 
MAST simulations.}
}
\end{center}
\end{figure}

\centori\ runs in parallel, with each MPI process advancing the physics
quantities in an allocated three-dimensional subdomain. Aggregation of quantities such as 
flux surface or volume integrals and averages are performed across appropriate sections 
of the process population via specially-written routines. Halo-swapping is necessarily frequent 
due to the extensive calculation of derivatives. For the small grid sizes that are suitable for 
MAST simulations, there is a tendency for the parallelisation to become communication-limited 
for relatively low process counts. There are, however, two different implementations of the key 
numerical routines available, which are optimised for different local grid sizes~\cite{Edwards}. 
Figure~\ref{fig:speedup} shows speed-up versus process count for simulations performed on 
HECToR at EPCC and HPC-FF at the J\"ulich Supercomputing Centre when a computational grid of 
$129 \times 65 \times 33$ and the ``eager'' implementation of the key routines, which perform 
better for a large number of processes, are used. The HECToR results were obtained using the 
Phase 2b system, 
based on Cray's XE6 hardware, which provides dual socket nodes with 2.1 GHz AMD Opteron 12-core 
processors and uses Cray's proprietary Gemini Interconnect; the PGI compiler was used. The
HPC-FF results were obtained using dual socket nodes with 2.93 GHz Intel Xeon X5570 quadcore 
processors and QDR Infiniband switch network; for these runs the Intel compiler was used.
In this particular case the application shows almost linear 
speed-up for process numbers of up to 128, and continues to display a significant speed-up even 
for 512 processes.  For a grid size of $129 \times 129 \times 129$  an almost linear speed-up is 
observed up to at least 512 processes (the largest number of processes used so far). Such a grid 
is larger than is normally used, but might be employed for ITER simulations in the future.
   
The computational domain is decomposed across the requested number of MPI
processes in a standard Cartesian communicator, with periodicity automatically
invoked in the two angular directions. Physics considerations suggest that the
number of grid intervals $N_{\psi}$, $N_{\theta}$ and $N_{\zeta}$ in the
radial, poloidal and toroidal directions should be roughly in
the ratio $4:2:1$ (a benchmarking study has confirmed that 
this aspect ratio delivers results that are close to optimal \cite{Edwards}).
In order to split each dimension into equal-sized portions across 
processes, the corresponding number of grid intervals must be one
greater than a power of two, and the number of processes in each dimension
must be an exact power of two. Thus, the computational grid used to model MAST
typically has $N_{\psi}= 129$, $N_{\theta} = 65$, $N_{\zeta} = 33$ over a
corresponding grid of $8\times 4\times 4$ processes.

Novel techniques are used to optimise the serial execution (on each parallel
process) of the numerical scheme within \centori. Specially designed derived
datatypes employing advanced pointer techniques are used, together with lazy
evaluation and the option to use strip-mining to tailor the code's
vectorisation. All the numerical vector operations (scalar and vector
products, gradient, divergence and curl derivatives), and many pure scalar or combined
scalar-vector operations, are contained within functional black boxes, hiding
the implementation details of the underlying datatypes and the potential
internal conversions between vector representations from the physics
programmer. This enables a researcher to convert a complicated physics
equation to a single line of code with ease. Full details are given
in~\cite{Edwards}. File handling is performed in parallel, with each process writing its own
output data files. However an option is under development that uses MPI-IO
routines to amalgamate the I/O more efficiently~\cite{Huhs}.

In contrast, the nature of the \grass\ two-dimensional equilibrium solution over the entire
poloidal cross-section of the plasma (and beyond) means that it is best
performed in the $(R,Z)$ laboratory coordinates on a single process, as is the
subsequent construction of the plasma coordinate system. This impacts only
weakly on the code performance, as it is not necessary to recalculate the
poloidal flux contours $\psi(R,Z)$ on timescales shorter than many
microseconds, and thus \grass\ is called only after many thousands of
evolution timesteps; $\Delta t \sim 10^{-9}$~s is the typical timestep used.
A typical run requires around 600 MB of RAM, and it takes around 12 hours
on 128 MPI processes to simulate 1$\,$ms of plasma evolution in typical tokamak 
conditions.


\subsection{Additional features within the \centori\ package}

The \centori\ code is best considered as a complete software package, rather than simply
a collection of source and input files. In addition to its normal role for
compilation, the makefile includes a number of utility functions that perform
tasks such as automatic generation of the code documentation, and the
creation of a tar file containing the entire source code, its documentation and
visualisation files, and the input and output files. This has proved to be of
great benefit in keeping all of the data from a given run together for archival
purposes.

The source code is self-documenting to a degree, using an included parser
program (\texttt{autodoc}) to generate html files for each subprogram from
specially-formatted comment lines within the code. In addition, a full
\LaTeX\/ manual is rigorously maintained to ensure its continued strict
agreement with the evolving source code (this paper is an abridged version of
the full manual).

A comprehensive visualisation suite has been developed to allow
straightforward interpretation of the physics output from \centori. The
program (\texttt{CentoriScope}) is written in the IDL
language~\cite{IDL}. Work is in progress to bring \centori\ into the
EU Integrated Tokamak Modelling (ITM~\cite{ITM}) framework.

The code is maintained within a private Subversion repository. Currently,
access to the code is obtainable only by prior permission from the
authors.

\section{Example outputs}
\label{sec:code_execution}

An early version of \centori\/ was used to study wave propagation in the vicinity of magnetic X-points; 
analytical results for the evolution of perturbed wave energy and plasma kinetic energy in the 
ideal MHD limit were recovered numerically \cite{McClements2004}. In this section we present two 
illustrative examples of calculations that can be performed using the full version of the code.

\subsection{Tearing mode in large aspect ratio tokamak plasma}

We demonstrate the capability of using \centori\/ to model MHD instabilities by considering the 
example of a tearing mode in a very large aspect ratio (minor radius $a = 0.36\,$m, major radius $R = 
16.8\,$m), circular cross-section tokamak plasma with a toroidal magnetic field of 9.7$\,$T. For this purpose 
we prescribe an equilibrium with uniform density ($10^{24}\,$m$^{-3}$) and temperature ($T_e=T_i=46\,$eV).
The resistivity, which we take to be given by the Spitzer expression [Eqs.~(\ref{eqn:spitzer}) 
and (\ref{eqn:tauce})], is then equal to $3.67\times 10^{-16}\,$s, and the Lundquist
number $S \equiv 4\pi av_A/(c^2\eta) \simeq 2\times 10^4$. The number of radial grid points (256)
was chosen to be sufficiently large that the resistive layer width $d \sim aS^{-2/5} \simeq 0.7\,$cm was 
well-resolved. The quantity $FF^{\prime}$, which is proportional to the toroidal current density in this 
large aspect ratio limit, was prescribed to have the following profile: 
\begin{equation}
FF^{\prime} = {FF^{\prime}(0)\over\left\{1+\hbox{sinh}^2\left[2.09\rho\right]\right\}^{3/2}},  
\end{equation}
where $FF^{\prime}(0)$ is a constant, chosen to ensure that the corresponding $q$-profile remained
above unity across the plasma, with $q=2$ at a normalised minor radius of about 0.65. This configuration 
is expected to be unstable to the growth of a tearing mode with poloidal and toroidal mode 
numbers $m=2$, $n=1$ \cite{FKR}. For the purpose of this calculation only the generalised Ohm's law and the 
ion momentum equation were 
used [Eqs. (\ref{eqn:momentum}) and (\ref{eqn:ohms_law})]. Ohm's law was reduced to the resistive MHD form, and 
viscosity was neglected in the momentum equation (except for the numerical viscosity described in 
Section~\ref{sec:vi_evolution}, which is required to suppress numerical oscillations, but is set at a level
which is sufficiently low for the system to be effectively inviscid). The electrostatic potential in this 
case was calculated not using Eq. (\ref{eqn:phistar2}) but by evolving the perpendicular ion velocity, 
identifying this as an 
${\bf E}\times{\bf B}$ drift, and integrating to obtain $\Phi$. Nonlinear terms were omitted from both 
Ohm's law and the momentum equation. Modes with $m/n$ equal to values of $q$ inside the plasma other than 2, 
such as the 3/2 mode, can also be unstable in the presence of a current density gradient; 
in general these modes cannot be exluded from simulations performed using a non-spectral code such 
as \centori\/. For this particular simulation, a Fourier filter was therefore applied at the end of each
time step to exclude all harmonics other than the dominant 2/1 mode.

\begin{figure}[ht]
\psfig{file=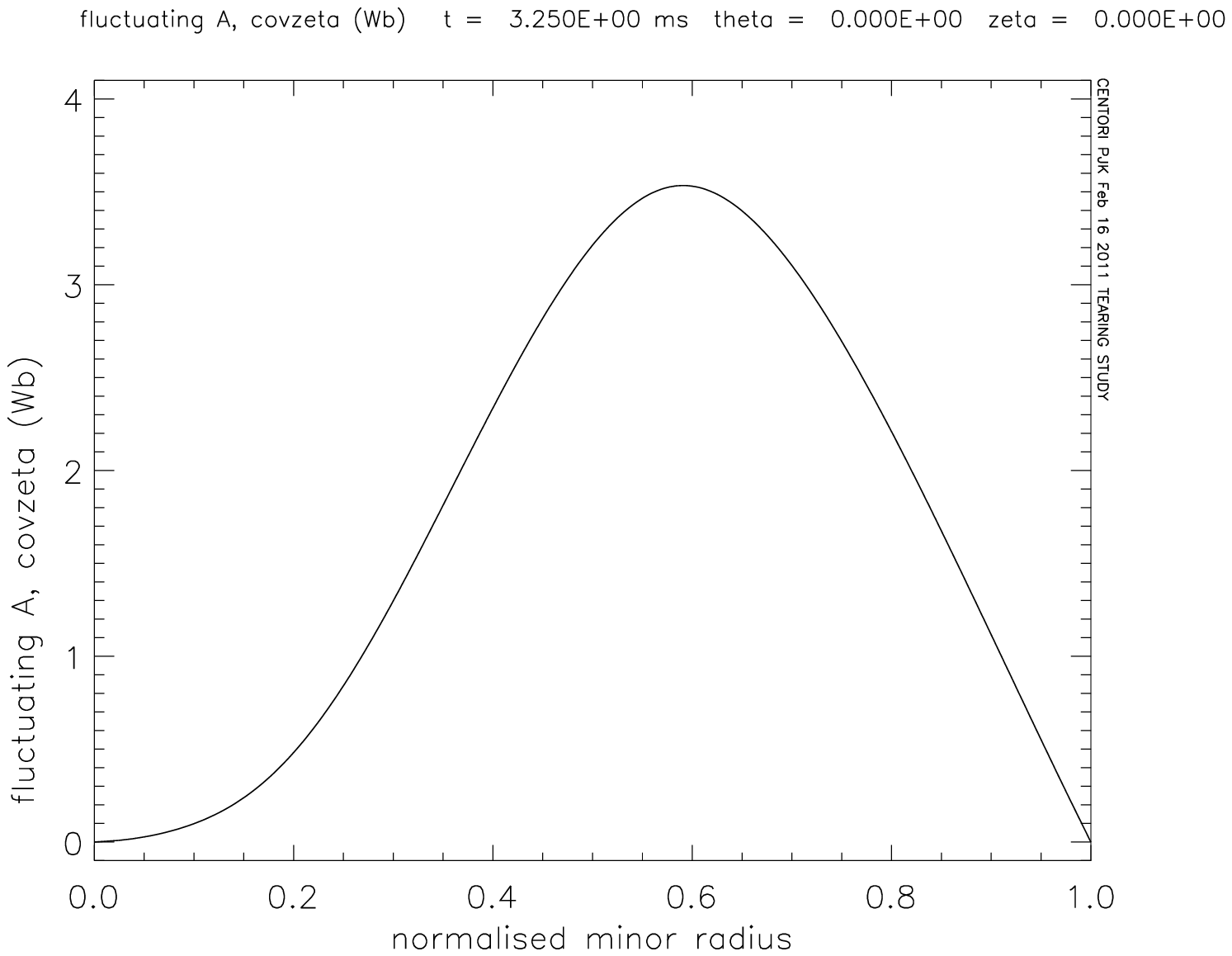,height=4.5cm,angle=0}
\psfig{file=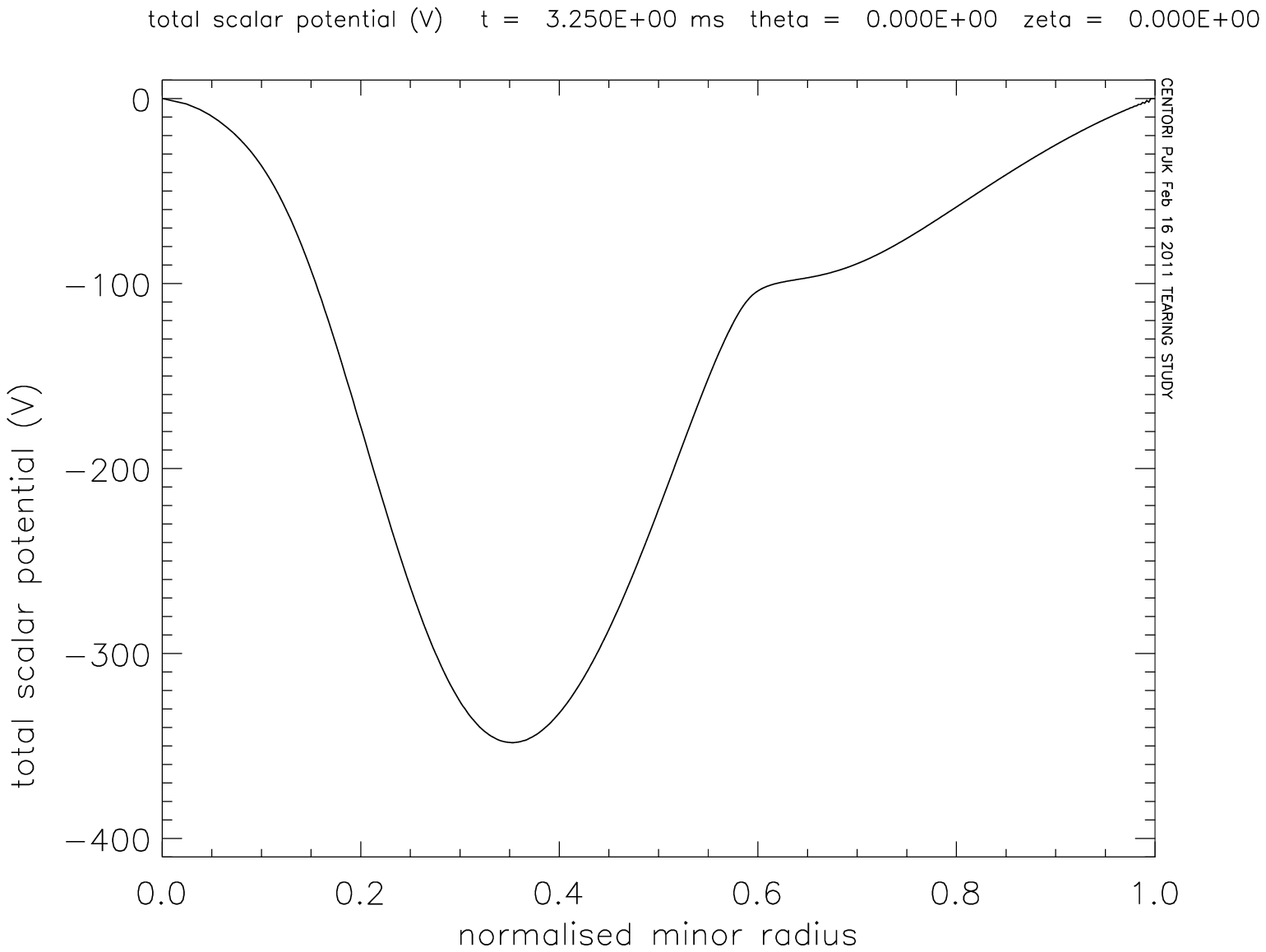,height=4.5cm,angle=0}
\parbox{14cm}{ \caption{\label{fig:Azeta}Radial profiles of $A_{\zeta}$ (left) and $\Phi$ (right)
in \centori\/ simulation of tearing mode in large aspect ratio tokamak.}
}
\end{figure}

As expected, the configuration described above was found using \centori\/ to be unstable to the 
growth of a 2/1 tearing mode. Figure \ref{fig:Azeta} shows snapshots of $A_{\zeta}$ and $\Phi$ during the 
tearing mode growth. The profiles of these quantities resemble those obtained using {\tt CUTIE} in a similar 
(although not identical) parameter regime; cf. Fig. 2 in Ref. \cite{Thyagaraja1994}, which was obtained with 
$S = 2 \times 10^4$ (defined in terms of the local resistivity at the magnetic axis) and relatively low 
viscosity (it should be noted that a precise comparison between tearing mode calculations performed using 
\centori\/ and the {\tt CUTIE} results presented in Ref. \cite{Thyagaraja1994} is not in fact
possible, since in the latter case a low aspect ratio ($R/a = 2.5$) was assumed for the purpose of calculating
the $q$-profile but the flux surfaces, as in all {\tt CUTIE} simulations, were taken to be concentric 
circles; in a toroidal code such as \centori\/ the circular flux surface limit can only be approached by taking 
the aspect ratio to be very large). The growth rate of the mode shown in Fig.~\ref{fig:Azeta} is approximately 
$2.5\times 10^{-3}/\tau_A$ where $\tau_A = a/v_A$ is the Alfv\'en time; this is comparable to the rate 
deduced analytically in Ref. \cite{FKR}, i.e. $\gamma\tau_A \sim S^{-3/5}$. It is somewhat lower than the 
rate found using {\tt CUTIE} in the low viscosity limit with $S = 2\times 10^4$ at the magnetic axis
($\gamma\tau_A \simeq 1.8\times 10^{-2}$) \cite{Thyagaraja1994}, but in this calculation the local resistivity
at the $q=2$ surface was higher than the value at the magnetic axis, implying a higher 2/1 tearing mode 
growth rate.      

\subsection{Turbulence simulation in conventional aspect ratio tokamak plasma}
\label{sec:turbsim}

\begin{figure}[ht]
\begin{center}
\psfig{file=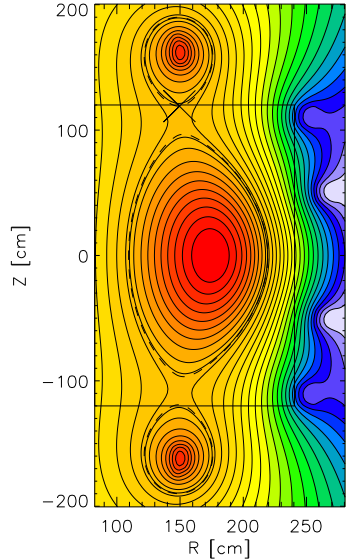,height=8cm,angle=0}
\parbox{14cm}{ \caption{\label{fig:equilibrium} Plot of $\psi$ contours for plasma equilibrium 
used in turbulence simulation described in Section 11. The inner rectangle indicates the 
plasma mask used to construct this equilibrium; note that the boundary of the mask lies 
outside the region of confined plasma, bounded by a thick black curve.}
}
\end{center}
\end{figure}

\begin{figure}[ht]
\begin{center}
\psfig{file=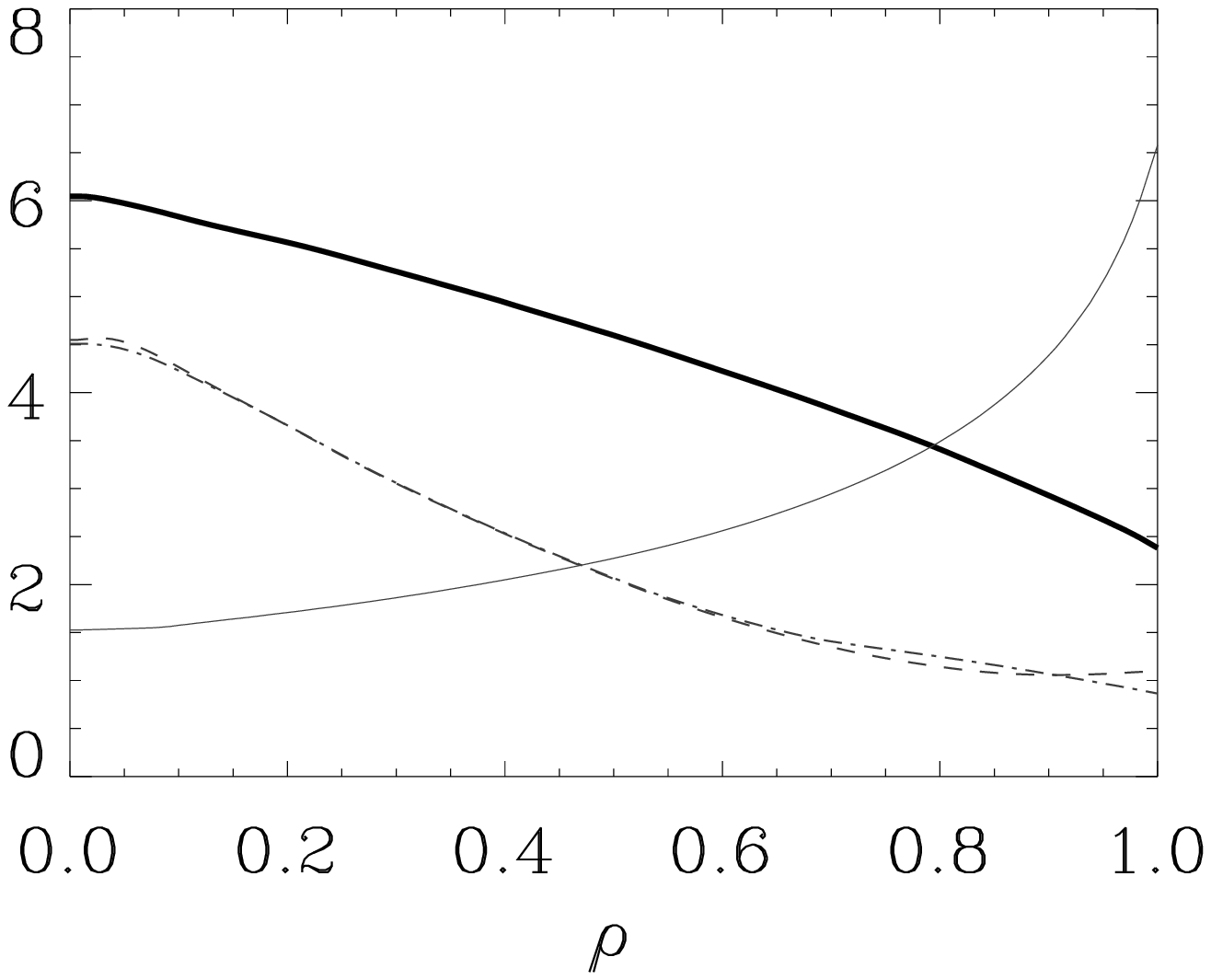,height=6cm,angle=0}
\parbox{14cm}{ \caption{\label{fig:profiles} Profiles used in turbulence simulation described 
in Section 11. The thick solid curve shows the electron density in units of $10^{19}$m$^{-3}$, 
the thin solid curve is the $q$-profile, and the dashed and dashed-dotted curves show respectively 
the electron temperature and ion temperature in keV.}
}
\end{center}
\end{figure}

We present here the results of a \centori\/ run simulating 1$\,$ms 
of a conventional aspect ratio tokamak plasma with minor radius 0.55$\,$m, major radius 1.67$\,$m, 
elongation 1.7 and triangularity 0.18; the equilibrium flux surface contours, computed using 
\grass\/, are shown in Fig.~\ref{fig:equilibrium}. 
The chosen plasma current was 1$\,$MA, the 
toroidal magnetic field 2.5$\,$T and the plasma volume 15$\,$m$^3$. The initial flux 
surface-averaged density, temperature and current density (primary quantities) were held 
approximately constant during the simulation by using adaptive sources of the form $S=
-\alpha(\langle f\rangle-\langle f_0\rangle)$, where $f$ is the primary variable profile,
$f_0$ is its initial profile, and $\alpha$ is an 
inverse reaction time response, set equal to the reciprocal of $\Delta t$, the \centori\/ time 
step (0.5$\,$ns, in this case). Both the initial ion velocity and the external momentum source 
$S_v$ were set equal to zero. The profiles of electron density, electron and ion temperatures, 
are shown in Fig.~\ref{fig:profiles}, together with the $q$-profile.  

\begin{figure}[ht]
\begin{center}
\psfig{file=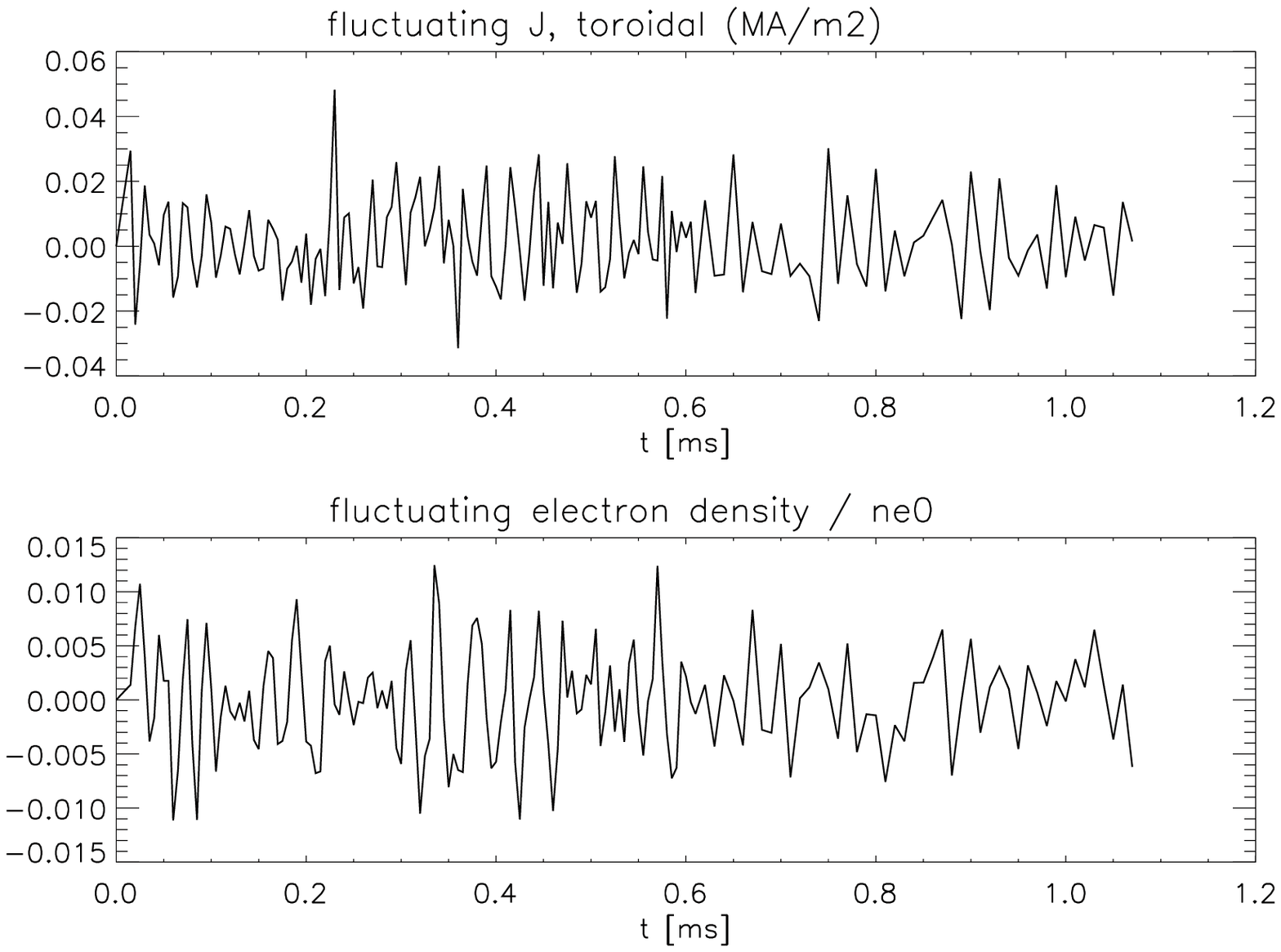,height=8cm,angle=0}
\parbox{14cm}{ \caption{\label{fig:fluctuations} Fluctuations in toroidal current density and 
electron density in outer midplane at $\rho = 0.46$.}
}
\end{center}
\end{figure}

The evolution of the sources follows that of the fluctuations; after an initial transient, lasting
around 100$\,\mu$s, they reached a quasi-steady level. The boundary conditions were those listed 
in Table 1. The spatial grid comprised 129 radial points, 65 poloidal points and 33 toroidal 
points. The run was executed on 64 processors of the HPC-FF machine at the J\"ulich Supercomputing
Centre, the total wall-clock time being approximately 18 hours. Figure~\ref{fig:fluctuations} 
shows the evolution of fluctuations in toroidal current density and electron density at 
$\rho = 0.46$, $\theta = 0$, $\zeta = 0$. It is evident from a comparison of the relative 
amplitudes of the temporal variations in these two quantities that the fluctuations are 
electromagnetic in character. 

\begin{figure}[ht]
\begin{center}
\psfig{file=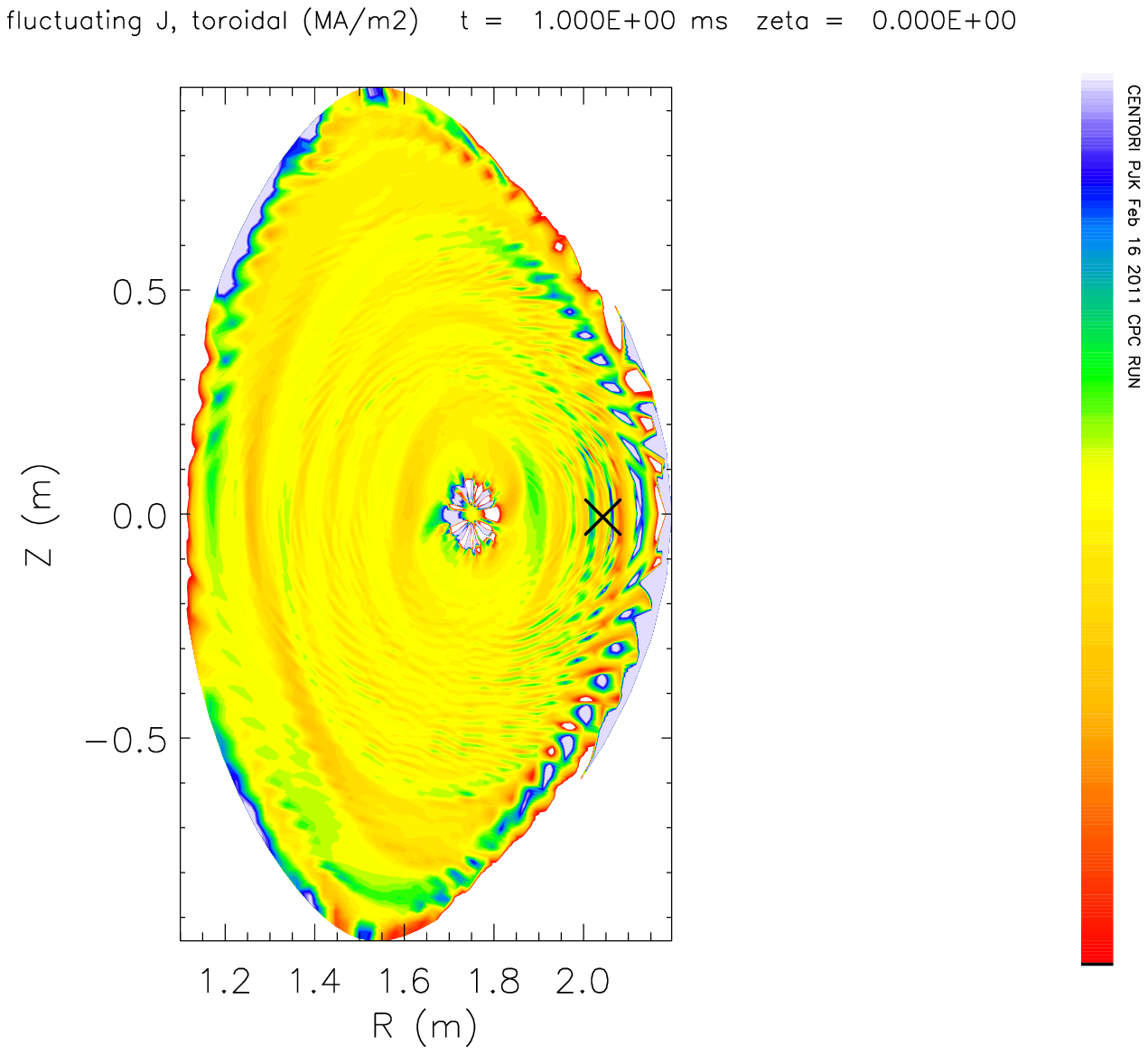,height=8cm,angle=0}
\parbox{14cm}{ \caption{\label{fig:contours} Contours of toroidal current density fluctuations in
poloidal plane at $t = 1\,$ms. The $\times$ symbol in the outer midplane indicates the approximate 
location corresponding to the results shown in Fig. 8.}
}
\end{center}
\end{figure}

\begin{figure}[ht]
\begin{center}
\psfig{file=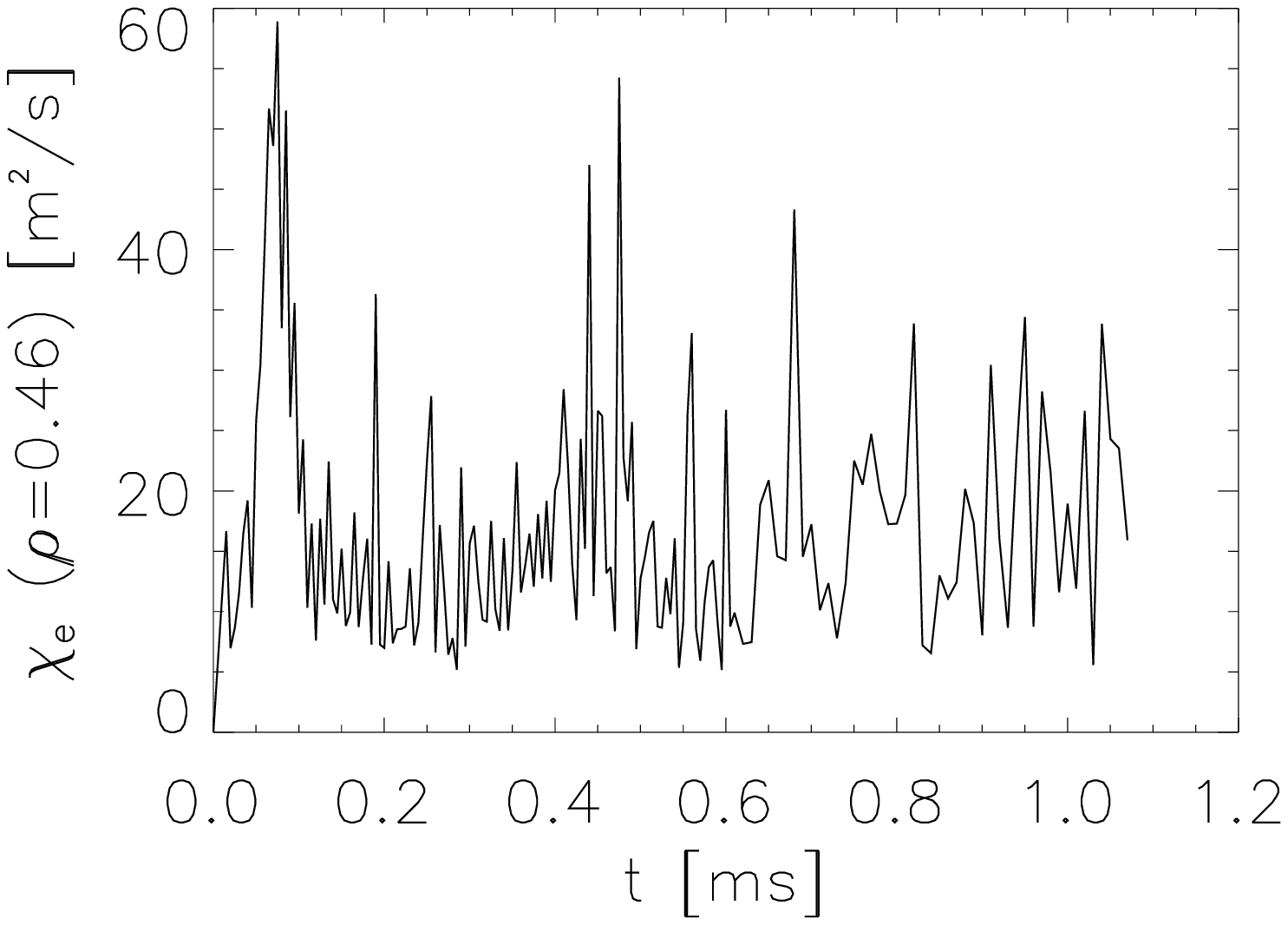,height=8cm,angle=0}
\parbox{14cm}{ \caption{\label{fig:chie} Evolution of electron thermal conductivity
in outer midplane at $\rho = 0.46$.}
}
\end{center}
\end{figure}

Figure~\ref{fig:contours} shows a poloidal cross-section of the toroidal current density 
fluctuations at $\zeta = 0$. The $\times$ symbol in this figure marks the location chosen for the 
sample of local fluctuations shown in Fig.~\ref{fig:fluctuations}. The temporal evolution of the 
electron thermal conductivity is shown in Fig.~\ref{fig:chie}. In the plasma turbulence 
literature this quantity is often normalised to the gyro-Bohm diffusivity $\chi_{\rm GB} = 
\rho_s^2c_s/L_T$ where $c_s = (T_e/m_i)^{1/2}$, $\rho_s = c_s/\omega_{ci}$ and $L_T = 
T_e/(dT_e/dr)$ is the temperature scale length \cite{Peeters}. In the case of the local plasma 
parameters corresponding to the results shown in Fig.~\ref{fig:chie}, $\rho_s^2c_s/L_T \simeq 
12\,$m$^2$s$^{-1}$; normalised to this value, the time-averaged thermal conductivity plotted in 
Fig.~\ref{fig:chie} is around 2, which is comparable to normalised $\chi_e$ values in 
gyro-kinetic simulations reported by Peeters and co-workers \cite{Peeters}. 

The results presented above can be used to estimate the magnitudes of the potential and inductive
contributions to the turbulent electric field; as noted in Section~\ref{sec:vi_evolution} 
only the potential electric field term is retained in the ion momentum equation in \centori. 
From Fig.~\ref{fig:fluctuations} we note that the electron density fluctuations have a relative 
amplitude $\tilde{n}_e/n_e$ of the order of 10$^{-2}$. Electron force balance implies that the associated 
electrostatic potential fluctuations $\tilde{\Phi}$ are of order $10^{-2}T_e/e \sim 20\,$V, since the 
electron temperature at this point in the plasma is about 2$\,$keV (cf. Fig.~\ref{fig:profiles}). 
Figure~\ref{fig:contours} indicates that the fluctuations have a characteristic scale length 
perpendicular to the magnetic field $L_{\perp}$ of order 10$^{-2}\,$m, suggesting potential electric 
field fluctuations of $\tilde{\Phi}/L_{\perp} \sim 2\,$kVm$^{-1}$. In contrast, the frequency ($\omega \sim
200\,$krad s$^{-1}$) and amplitude ($\sim 0.03\,$MAm$^{-2}$) of the current fluctuations $\tilde{J}$ shown 
in the upper frame of Fig.~\ref{fig:fluctuations} imply inductive electric fields of order 
$\omega\mu_0\tilde{J}L_{\perp}^2 \sim 1\,$Vm$^{-1}$ ($\mu_0$ being the permeability of free space). Thus, 
for the parameters of this simulation (which are fairly representative of hot tokamak plasmas), the 
potential component of the fluctuating electric field is around three orders of magnitude larger than 
the inductive component, and our neglect of the latter in the ion momentum equation is therefore fully 
justified.  

\section{Conclusion}
\label{sec:conclusion}

We have presented a comprehensive description of a novel two-fluid electromagnetic plasma 
turbulence code, \centori\/, together with sample output from a \centori\/ simulation 
of a large aspect ratio tokamak plasma. The code is used to compute self-consistently the time 
evolution of plasma fluid quantities and fields in a toroidal configuration of arbitrary aspect 
ratio and plasma beta. The code is parallelised, and the equations are represented in fully 
finite difference form, ensuring good scalability. The equations are solved in a plasma coordinate system
that is defined such that the Jacobian of the transformation from laboratory coordinates is a function only
of the equilibrium poloidal flux, thereby accelerating vector operations and the evaluation 
of flux surface averages. \grass, a subroutine of \centori, is used to determine the plasma equilibrium 
(and hence the plasma coordinates in which the fluid and Maxwell equations are solved) by computing 
the steady-state solutions of a diffusion equation with a pseudo-time derivative. The physics model 
implemented in \centori\/ is based solidly on that used in the highly-successful {\tt CUTIE} code,  
and we are confident that it will prove to be a powerful tool 
for the study of heat, particle and momentum transport in tokamak plasmas. In a forthcoming paper
we will report the results of the first global simulations, performed using \centori\/, of electromagnetic, 
nonlinearly-saturated turbulence and transport in a spherical tokamak plasma (MAST).

\section*{Acknowledgements}

This work was supported by EPSRC grants EP/I501045 (as part of the RCUK Energy Programme),
EP/H00212X/1 and EP/H002081/1, and the European Communities under the Contract of Association 
between EURATOM and CCFE. The views and opinions expressed herein do not necessarily reflect those
of the European Commission. We would like to thank Dr F. Militello (CCFE) and two anonymous 
referees for helpful suggestions that have led to improvements in this paper.

\end{document}